\ifdefined\newapj
\documentclass[twocolumn]{aastex62} 
\else
\documentclass[chicago]{emulateapj} 
\fi

\usepackage{amsmath}
\usepackage{natbib}
\usepackage{graphicx}
\usepackage{epsf}
\usepackage{subfigure}
\usepackage{color}
\usepackage{threeparttable}
\usepackage{comment}
\usepackage{epsfig}
\usepackage{xspace}
\usepackage{enumitem}
\usepackage{hyperref}
\usepackage{ulem}
\usepackage[usenames,dvipsnames,svgnames]{xcolor}
\usepackage{adjustbox}

\DeclareGraphicsExtensions{.jpg,.pdf,.pdf,.eps,.ps}

\newcommand{\refsec}[1]{Section~\ref{sec:#1}}
\newcommand{\refeq}[1]{Eq.~(\ref{eq:#1})}
\newcommand{\refssec}[1]{Section~\ref{ssec:#1}}
\newcommand{\reffig}[1]{Figure~\ref{fig:#1}}

\newcommand{\reftab}[1]{Table~\ref{tab:#1}}

\def\beq{\begin{equation}}
\def\eeq{\end{equation}}
\def\beqn{\begin{align}}
\def\eeqn{\end{align}}
\def\bl{\pmb{\ell}}
\def\bL{\mathbf{L}}
 
\def\cLppHat{\ensuremath{C^{\hat{\phi} \hat{\phi}}_{L}}}
\def\hatcLpp{\ensuremath{\hat{C}^{\phi \phi}_{L}}}

\newcommand{\sptpol}{SPTpol\xspace}
\newcommand{\planck}{\textit{Planck}\xspace}

\newcommand{\LCDM}{\mbox{$\Lambda$CDM}\xspace}

\newcommand{\mvSysCal}{\ensuremath{0.023}}
\newcommand{\mvSysBeam}{\ensuremath{0.008}}
\newcommand{\mvSysFg}{\ensuremath{0.004}}

\newcommand{\ppSysCal}{\ensuremath{0.039}}
\newcommand{\ppSysBeam}{\ensuremath{0.010}}

\newcommand{\ttSysCal}{\ensuremath{0.008}}
\newcommand{\ttSysBeam}{\ensuremath{0.005}}
\newcommand{\ttSysFg}{\ensuremath{0.008}}

\newcommand{\mvAmp}{\ensuremath{0.944}}
\newcommand{\mvAmpStat}{\ensuremath{0.058}}
\newcommand{\mvAmpSys}{\ensuremath{0.025}}

\newcommand{\ppAmp}{\ensuremath{0.906}}
\newcommand{\ppAmpStat}{\ensuremath{0.090}}
\newcommand{\ppAmpSys}{\ensuremath{0.040}}

\newcommand{\ttAmp}{\ensuremath{0.835}}
\newcommand{\ttAmpStat}{\ensuremath{0.103}}
\newcommand{\ttAmpSys}{\ensuremath{0.012}}

\newcommand{\mvNullSig}{\ensuremath{39\,\sigma}}    
\newcommand{\mvPercent}{\ensuremath{ 6 \%}}      
\newcommand{\ppLensSig}{\ensuremath{10.1\,\sigma}}        
\newcommand{\ppPercent}{\ensuremath{10 \%}}      
\newcommand{\ttPercent}{\ensuremath{12 \%}}      

\newcommand{\mvTotPercent}{\ensuremath{ 7 \%}}      

\def\mv{MV}
\def\pol{POL}
\def\pp{POL}
\def\tt{T}
\def\tcal{\ensuremath{T_{\rm cal}}}
\def\pcal{\ensuremath{P_{\rm cal}}}
\def\muKsq{$\mu$K$^2$}
\def\nzero{$N_L^{(0)}$}
\def\rdn{\ensuremath{N_L^{(0), \rm RD}}}
\def\none{\ensuremath{N_L^{(1)}}}

\def\cmbLppHat{\ensuremath{C^{\hat{\phi} \hat{\phi}}_{L}}}

\definecolor{orange}{HTML}{BF482F}

\begin{document}

\ifdefined\newapj
\title{A Measurement of the Cosmic Microwave Background Lensing Potential and Power Spectrum from 500 deg$^2$ of SPTpol Temperature and Polarization Data}
\else
\title{A measurement of the cosmic microwave background lensing potential and power spectrum from 500 deg$^2$ of SPTpol temperature and polarization data}
\fi

\ifdefined\newapj
\input lens500d_authors_apj.tex 
\else
\def\KICPChicago{1}
\def\AAUChicago{2}
\def\UIO{3}
\def\Cardiff{4}
\def\FNAL{5}
\def\NIST{6}
\def\Berkeley{7}
\def\ArgonneHEP{8}
\def\Melbourne{9}
\def\PhysicsUChicago{10}
\def\EFIChicago{11}
\def\McGill{12}
\def\UKZN{13}
\def\TAPIRCaltech{14}
\def\Caltech{15}
\def\LBNL{16}
\def\CIFAR{17}
\def\ColoradoAPS{18}
\def\HarveyMudd{19}
\def\esogarching{20}
\def\ColoradoPhys{21}
\def\illast{22}
\def\illphy{23}
\def\UChicago{24}
\def\SLAC{25}
\def\Stanford{26}
\def\Davis{27}
\def\IAP{28}
\def\Michigan{29}
\def\ArgonneMSD{30}
\def\KIPAC{31}
\def\Minnesota{32}
\def\CaseWestern{33}
\def\Yale{34}
\def\ArtInstChicago{35}
\def\ThreeSpeedLogic{36}
\def\CfA{37}
\def\Dunlap{38}
\def\UToronto{39}
\def\Maryland{40}
\def\UCLA{41}
\def\kwemail{$\dagger$}
\def\mmemail{$\ddagger$}

\email{$\dagger$ wlwu@kicp.uchicago.edu}
\email{$\ddagger$ l.m.mocanu@astro.uio.no \\}

\shortauthors{W.~L.~K.~Wu, L.~M.~Mocanu, et al.}
\author{
  W.~L.~K.~Wu\altaffilmark{\KICPChicago,\kwemail},
  L.~M.~Mocanu\altaffilmark{\KICPChicago,\AAUChicago,\UIO,\mmemail},
  P.~A.~R.~Ade\altaffilmark{\Cardiff},
  A.~J.~Anderson\altaffilmark{\FNAL},
  J.~E.~Austermann\altaffilmark{\NIST},
  J.~S.~Avva\altaffilmark{\Berkeley},
  J.~A.~Beall\altaffilmark{\NIST},
  A.~N.~Bender\altaffilmark{\KICPChicago,\ArgonneHEP},
  B.~A.~Benson\altaffilmark{\KICPChicago,\AAUChicago,\FNAL},
  F.~Bianchini\altaffilmark{\Melbourne},
  L.~E.~Bleem\altaffilmark{\KICPChicago,\ArgonneHEP},
  J.~E.~Carlstrom\altaffilmark{\KICPChicago,\AAUChicago,\ArgonneHEP,\PhysicsUChicago,\EFIChicago},
  C.~L.~Chang\altaffilmark{\KICPChicago,\AAUChicago,\ArgonneHEP},
  H.~C.~Chiang\altaffilmark{\McGill,\UKZN},
  R.~Citron\altaffilmark{\KICPChicago},
  C.~Corbett~Moran\altaffilmark{\TAPIRCaltech},
  T.~M.~Crawford\altaffilmark{\KICPChicago,\AAUChicago},
  A.~T.~Crites\altaffilmark{\KICPChicago,\AAUChicago,\Caltech},
  T.~de~Haan\altaffilmark{\Berkeley,\LBNL},
  M.~A.~Dobbs\altaffilmark{\McGill,\CIFAR},
  W.~Everett\altaffilmark{\ColoradoAPS},
  J.~Gallicchio\altaffilmark{\KICPChicago,\HarveyMudd},
  E.~M.~George\altaffilmark{\Berkeley,\esogarching},
  A.~Gilbert\altaffilmark{\McGill},
  N.~Gupta\altaffilmark{\Melbourne},
  N.~W.~Halverson\altaffilmark{\ColoradoAPS,\ColoradoPhys},
  N.~Harrington\altaffilmark{\Berkeley},
  J.~W.~Henning\altaffilmark{\KICPChicago,\ArgonneHEP},
  G.~C.~Hilton\altaffilmark{\NIST},
  G.~P.~Holder\altaffilmark{\CIFAR,\illast,\illphy},
  W.~L.~Holzapfel\altaffilmark{\Berkeley},
  Z.~Hou\altaffilmark{\KICPChicago},
  J.~D.~Hrubes\altaffilmark{\UChicago},
  N.~Huang\altaffilmark{\Berkeley},
  J.~Hubmayr\altaffilmark{\NIST},
  K.~D.~Irwin\altaffilmark{\SLAC,\Stanford},
  L.~Knox\altaffilmark{\Davis},
  A.~T.~Lee\altaffilmark{\Berkeley,\LBNL},
  D.~Li\altaffilmark{\NIST,\SLAC},
  A.~Lowitz\altaffilmark{\KICPChicago},
  A.~Manzotti\altaffilmark{\KICPChicago,\IAP},
  J.~J.~McMahon\altaffilmark{\Michigan},
  S.~S.~Meyer\altaffilmark{\KICPChicago,\AAUChicago,\PhysicsUChicago,\EFIChicago},
  M.~Millea\altaffilmark{\Berkeley},
  J.~Montgomery\altaffilmark{\McGill,\CIFAR},
  A.~Nadolski\altaffilmark{\illast,\illphy},
  T.~Natoli\altaffilmark{\KICPChicago,\AAUChicago},
  J.~P.~Nibarger\altaffilmark{\NIST},
  G.~I.~Noble\altaffilmark{\McGill},
  V.~Novosad\altaffilmark{\ArgonneMSD},
  Y.~Omori\altaffilmark{\Stanford,\KIPAC},
  S.~Padin\altaffilmark{\KICPChicago,\AAUChicago,\Caltech},
  S.~Patil\altaffilmark{\Melbourne},
  C.~Pryke\altaffilmark{\Minnesota},
  C.~L.~Reichardt\altaffilmark{\Melbourne},
  J.~E.~Ruhl\altaffilmark{\CaseWestern},
  B.~R.~Saliwanchik\altaffilmark{\UKZN,\Yale},
  J.T.~Sayre\altaffilmark{\ColoradoAPS,\ColoradoPhys},
  K.~K.~Schaffer\altaffilmark{\KICPChicago,\EFIChicago,\ArtInstChicago},
  C.~Sievers\altaffilmark{\KICPChicago},
  G.~Simard\altaffilmark{\McGill},
  G.~Smecher\altaffilmark{\McGill,\ThreeSpeedLogic},
  A.~A.~Stark\altaffilmark{\CfA},
  K.~T.~Story\altaffilmark{\Stanford,\KIPAC},
  C.~Tucker\altaffilmark{\Cardiff},
  K.~Vanderlinde\altaffilmark{\Dunlap,\UToronto},
  T.~Veach\altaffilmark{\Maryland},
  J.~D.~Vieira\altaffilmark{\illast,\illphy},
  G.~Wang\altaffilmark{\ArgonneHEP},
  N.~Whitehorn\altaffilmark{\UCLA},
  and
  V.~Yefremenko\altaffilmark{\ArgonneHEP}
}

\altaffiltext{\KICPChicago}{Kavli Institute for Cosmological Physics, University of Chicago, 5640 South Ellis Avenue, Chicago, IL, USA 60637}
\altaffiltext{\AAUChicago}{Department of Astronomy and Astrophysics, University of Chicago, 5640 South Ellis Avenue, Chicago, IL, USA 60637}
\altaffiltext{\UIO}{Institute of Theoretical Astrophysics, University of Oslo, P.O.Box 1029 Blindern, N-0315 Oslo, Norway}
\altaffiltext{\Cardiff}{Cardiff University, Cardiff CF10 3XQ, United Kingdom}
\altaffiltext{\FNAL}{Fermi National Accelerator Laboratory, MS209, P.O. Box 500, Batavia, IL 60510}
\altaffiltext{\NIST}{NIST Quantum Devices Group, 325 Broadway Mailcode 817.03, Boulder, CO, USA 80305}
\altaffiltext{\ArgonneHEP}{High Energy Physics Division, Argonne National Laboratory, 9700 S. Cass Avenue, Argonne, IL, USA 60439}
\altaffiltext{\Melbourne}{School of Physics, University of Melbourne, Parkville, VIC 3010, Australia}
\altaffiltext{\PhysicsUChicago}{Department of Physics, University of Chicago, 5640 South Ellis Avenue, Chicago, IL, USA 60637}
\altaffiltext{\EFIChicago}{Enrico Fermi Institute, University of Chicago, 5640 South Ellis Avenue, Chicago, IL, USA 60637}
\altaffiltext{\UKZN}{School of Mathematics, Statistics \& Computer Science, University of KwaZulu-Natal, Durban, South Africa}
\altaffiltext{\SLAC}{SLAC National Accelerator Laboratory, 2575 Sand Hill Road, Menlo Park, CA 94025}
\altaffiltext{\TAPIRCaltech}{TAPIR, Walter Burke Institute for Theoretical Physics, California Institute of Technology, 1200 E California Blvd, Pasadena, CA, USA 91125}
\altaffiltext{\Caltech}{California Institute of Technology, MS 249-17, 1216 E. California Blvd., Pasadena, CA, USA 91125}
\altaffiltext{\Berkeley}{Department of Physics, University of California, Berkeley, CA, USA 94720}
\altaffiltext{\LBNL}{Physics Division, Lawrence Berkeley National Laboratory, Berkeley, CA, USA 94720}
\altaffiltext{\McGill}{Department of Physics, McGill University, 3600 Rue University, Montreal, Quebec H3A 2T8, Canada}
\altaffiltext{\CIFAR}{Canadian Institute for Advanced Research, CIFAR Program in Cosmology and Gravity, Toronto, ON, M5G 1Z8, Canada}
\altaffiltext{\ColoradoAPS}{Department of Astrophysical and Planetary Sciences, University of Colorado, Boulder, CO, USA 80309}
\altaffiltext{\HarveyMudd}{Harvey Mudd College, 301 Platt Blvd., Claremont, CA 91711}
\altaffiltext{\esogarching}{European Southern Observatory, Karl-Schwarzschild-Str. 2, 85748 Garching bei M\"{u}nchen, Germany}
\altaffiltext{\ColoradoPhys}{Department of Physics, University of Colorado, Boulder, CO, USA 80309}
\altaffiltext{\illast}{Astronomy Department, University of Illinois at Urbana-Champaign, 1002 W. Green Street, Urbana, IL, USA 61801}
\altaffiltext{\illphy}{Department of Physics, University of Illinois Urbana-Champaign, 1110 W. Green Street, Urbana, IL, USA 61801}
\altaffiltext{\UChicago}{University of Chicago, 5640 South Ellis Avenue, Chicago, IL, USA 60637}
\altaffiltext{\Stanford}{Dept. of Physics, Stanford University, 382 Via Pueblo Mall, Stanford, CA 94305}
\altaffiltext{\Davis}{Department of Physics, University of California, One Shields Avenue, Davis, CA, USA 95616}
\altaffiltext{\IAP}{Institut d'Astrophysique de Paris, 98 bis boulevard Arago, 75014 Paris, France}
\altaffiltext{\Michigan}{Department of Physics, University of Michigan, 450 Church Street, Ann  Arbor, MI, USA 48109}
\altaffiltext{\ArgonneMSD}{Materials Sciences Division, Argonne National Laboratory, 9700 S. Cass Avenue, Argonne, IL, USA 60439}
\altaffiltext{\KIPAC}{Kavli Institute for Particle Astrophysics and Cosmology, Stanford University, 452 Lomita Mall, Stanford, CA 94305}
\altaffiltext{\Minnesota}{School of Physics and Astronomy, University of Minnesota, 116 Church Street S.E. Minneapolis, MN, USA 55455}
\altaffiltext{\UCLA}{Department of Physics and Astronomy, University of California, Los Angeles, CA, USA 90095}
\altaffiltext{\CaseWestern}{Physics Department, Center for Education and Research in Cosmology and Astrophysics, Case Western Reserve University, Cleveland, OH, USA 44106}
\altaffiltext{\Yale}{Department of Physics, Yale University, 217 Prospect Street, New Haven, CT, USA 06511}
\altaffiltext{\ArtInstChicago}{Liberal Arts Department, School of the Art Institute of Chicago, 112 S Michigan Ave, Chicago, IL, USA 60603}
\altaffiltext{\ThreeSpeedLogic}{Three-Speed Logic, Inc., Vancouver, B.C., V6A 2J8, Canada}
\altaffiltext{\CfA}{Harvard-Smithsonian Center for Astrophysics, 60 Garden Street, Cambridge, MA, USA 02138}
\altaffiltext{\Dunlap}{Dunlap Institute for Astronomy \& Astrophysics, University of Toronto, 50 St George St, Toronto, ON, M5S 3H4, Canada}
\altaffiltext{\UToronto}{Department of Astronomy \& Astrophysics, University of Toronto, 50 St George St, Toronto, ON, M5S 3H4, Canada}
\altaffiltext{\Maryland}{Department of Astronomy, University of Maryland College Park, MD, USA 20742}

\fi

\begin{abstract}
We present a measurement of the cosmic microwave background (CMB) lensing potential using 500\,deg$^2$ of 150 GHz data from the \sptpol\ receiver on the South Pole Telescope.
The lensing potential is reconstructed with signal-to-noise per mode greater than unity at lensing multipoles $L \lesssim 250$,
using a quadratic estimator on a combination of CMB temperature and polarization maps. 
We report measurements of the lensing potential power spectrum in the multipole range of $100< L < 2000$ from sets of temperature-only (\tt), polarization-only (\pp), and minimum-variance (\mv) estimators.
We measure the lensing amplitude by taking 
the ratio of the measured spectrum to the expected spectrum from the best-fit \LCDM model to the \planck 2015 TT+lowP+lensing dataset.
For the minimum-variance estimator, we find
$A_{\rm{\mv}} = \mvAmp \pm \mvAmpStat{\rm \,(Stat.)}\pm\mvAmpSys{\rm\,(Sys.)}$;
restricting to only polarization data, we find $A_{\rm{\pp}} = \ppAmp \pm \ppAmpStat {\rm\,(Stat.)} \pm \ppAmpSys {\rm\,(Sys.)}$.  
Considering statistical uncertainties alone, this is the most precise polarization-only lensing amplitude constraint to date (\ppLensSig), 
and is more precise than our temperature-only constraint. 
We perform null tests and consistency checks and find no evidence for significant contamination. 

\end{abstract}


\section{Introduction}
\label{sec:intro}
\setcounter{footnote}{0}
Gravitational potentials associated with large-scale structure deflect the paths of cosmic
microwave background (CMB) photons as they propagate from the surface of
last scattering -- a process called gravitational lensing~\citep{blanchard87}. 
Gravitational lensing breaks the statistical isotropy of the CMB and introduces correlations 
across CMB temperature and polarization fluctuations on different angular scales.
These correlations are proportional to the projected gravitational potentials integrated 
along the line of sight and therefore 
can be used to reconstruct the lensing potential~\citep{lewis06}.
The lensing potential is a probe of the growth of large-scale structure
and the geometry of the universe between the epoch of recombination and today.
Thus, from CMB observations alone, 
we can extract information about the universe 
at both the redshift of last scattering ($z \simeq 1100$) and 
the redshifts of structure formation  ($z \lesssim 3$) and dark energy domination. 
This makes CMB lensing a powerful tool for pursuing some of the most ambitious goals in cosmology and
particle physics today~\citep[e.g.][]{cmbs4-sb1}, including constraining
the sum of neutrino masses and the amplitude of primordial gravitational 
waves~\citep{lesgourgues06b, kamionkowski16}.

While using the CMB lensing measurements from this work, we will not detect the sum of neutrino masses 
or significantly constrain 
primordial gravitational waves through {\it delensing}~\citep{manzotti17},
these measurements will nevertheless provide
relevant constraints to parameters of the
standard cosmological model \LCDM. 
This is of particular interest currently because some optical probes of gravitational lensing are in mild tension
with \planck's CMB+lensing constraints on matter density and fluctuations~\citep{des18_3x2, kids18, chiaki19}.
The optical lensing measurements use the effect of gravitational
lensing on the apparent shapes of background galaxies to measure the intervening gravitational potentials. 
Compared with using galaxies, an attractive characteristic of CMB lensing is that the source plane has nearly Gaussian statistics 
with a well-characterized angular power spectrum
and is at a high and well-known redshift $z=1089.8 \pm 0.2$~\citep{planck15-11, planck18-6}.
Owing to the high-redshift source plane, 
CMB lensing probes the integrated matter fluctuations to redshifts beyond
optical surveys. 
Furthermore, CMB lensing measurements have different instrumental and astrophysical systematics
compared with those from optical surveys.
The use of independent probes will therefore help us investigate the source of the tension. 
It is thus an important goal for us  to 
identify potential sources of instrumental and/or astrophysical systematics
as the field advances the precision of CMB lensing measurements. 

In the last few years, CMB lensing has entered the era of precision measurements. 
This lensing effect was first detected through cross correlations with radio sources~\citep{smith07a}.
Since then, the CMB lensing potential power spectrum has been measured by multiple
groups using temperature data only (\tt, \citealt{das11, vanengelen12, planck13-17, omori17}), 
polarization data only (\pp, \citealt{polarbear2014a, bicep2keck16}), and combinations of temperature and 
polarization data~\citep{story14, sherwin16, planck18-8}.
The most precise lensing amplitude measurement, at 40\,$\sigma$, comes from 
\planck's minimum-variance (\mv) estimator that combines both temperature 
and polarization estimators; 
in that measurement, the temperature reconstruction contributes most
of the signal-to-noise ratio (S/N).
More generally, in prior lensing measurements that used both temperature and polarization maps,
the \tt\ estimator has always dominated the overall measurement precision. 
In this work, for the first time,
the \pp\ measurement is more constraining than the \tt\ measurement.
Furthermore, if we consider only the statistical uncertainty, 
we have a \ppPercent\ constraint (\ppLensSig) on the lensing amplitude 
using polarization data alone -- the tightest constraint of its kind to date. 

In this work, we extend the lensing measurement to a 500\,deg$^2$ field 
from the 100\,deg$^2$ field of~\citet[][hereafter S15]{story14}. 
We build on the lensing pipeline presented in S15 
with two main modifications.
First, instead of including the Monte Carlo (MC) bias -- the difference between 
the recovered lensing spectrum from simulation and 
the input spectrum -- as a systematic uncertainty, 
we identify its main contributor and correct for the bias using a multiplicative
correction factor.
Second, instead of treating extragalactic foreground biases as negligible, 
we subtract an expected foreground bias from the \tt\ and \mv\ lensing spectra. 

Because the input CMB maps are of similar depths as those used in S15,
we also measure lensing modes with S/N better than unity for $L \lesssim 250$. 
However, in this analysis, we have $\sim$5 times more sky area and therefore are 
able to make a more precise measurement of the lensing spectrum.
With both the \tt\ and \pp\ lensing amplitudes well constrained,
we improve the precision of the \mv\ lensing amplitude measurement 
from S15's 14\% to \mvPercent. 
This is approaching the $\sim$3\% precision of the lensing amplitude measurement
from \planck~\citep{planck18-8}. 
These two measurements arrive at their respective precisions from
very different regimes: 
while the \planck measurement covers 67\% of the sky, each lensing mode is measured
with low S/N; 
our measurement covers only 1\% of the sky, but many lensing modes are measured with 
high S/N. 
When our measurements are projected to cosmological parameter space, 
the constraint in the $\sigma_8-\Omega_M$ plane is only slightly weaker than those
from \planck~\citep{planck18-8}. 
This will be useful in illuminating the aforementioned mild tension between
optical probes and \planck. 
In a future paper, we will present cosmological parameter constraints and
comparisons with other lensing probes. 

This paper is organized as follows: in~\refsec{data_sim}, we describe the dataset
used and the simulated skies generated for this analysis. In~\refsec{lensing}, we 
summarize the lensing analysis pipeline and describe 
new aspects. We present the lensing measurements in~\refsec{results} and
show that our measurements are robust against systematics in~\refssec{null}.
In~\refssec{sys}, we account for systematic uncertainties from sources 
that can bias the lensing measurement.
We conclude in~\refsec{conclusion}. 

\section{Data and Simulations}
\label{sec:data_sim}
In this section, we describe the \sptpol\ survey, the data processing steps taken
to generate the data maps, and the simulated skies generated for this analysis.

\subsection{\sptpol 500\,deg$^2$  survey}
\label{ssec:survey}
The South Pole Telescope \citep[SPT,][]{padin08, carlstrom11} is a 10-meter diameter 
off-axis Gregorian telescope located at the Amundsen-Scott South Pole Station 
in Antarctica. 
In this work, we use data from the 150\,GHz detectors from 
the first polarization-sensitive receiver on SPT, \sptpol. 
The \sptpol 500\,deg$^2$ survey
spans 4 hours in right ascension (R.A.), from $22^h$ to $2^h$, and 
15 degrees in declination (dec.), from $-65\deg$ to $-50\deg$.
We use observations conducted between 
April 30, 2013 and October 27, 2015 (after the 100\,deg$^2$ 
survey for S15 was finished),
which include 3491 independent maps
of the 500\,deg$^2$ survey footprint. 
The field was observed using two strategies. 
Initially, we used a ``lead-trail'' scanning strategy, 
where the field was divided into two halves in R.A..
The telescope first scanned the lead half, and then switched 
to the trail half such that they were both observed over the same range in azimuth, 
and therefore the same patch of ground. 
In May 2014, the scanning strategy was switched to a ``full-field" strategy, 
in which the full R.A. range of the field was covered in a single observation.

\subsection{Data processing}
\label{ssec:data}
The data reduction for this set of maps follows that applied to
the TE/EE power spectrum analysis of the same field~\citep[][hereafter H18]{henning18}. 
We therefore highlight here only aspects that are different or are particularly relevant for this analysis. 

An observation is built from a collection of constant-elevation scans, where the
telescope moves from one end of the R.A. range to the other.
For every scan,
we filter the time streams, which corresponds in map space to mode removal along
the scan direction.
We subtract from each detector's time stream a Legendre polynomial 
as an effective high-pass filter. 
We choose a 3rd order polynomial for the lead-trail scans and 
a 5th order polynomial for the full-field scans. 
We combine this with a high-pass filter with a cutoff frequency corresponding to 
angular multipole $\ell =100$ in the scan direction.
During the polynomial and high-pass filtering step, 
regions within $5'$ of point sources brighter than 50\,mJy at 150\,GHz are masked 
in the time streams to avoid ringing artifacts.
Finally, we low-pass the time streams at an effective multipole of $\ell=7500$ in the scan direction
to avoid high-frequency noise 
aliasing into the signal due to the adopted pixel resolution of $1^{\prime}$ in the maps. 

Electrical cross talk between detectors can bias our measurement.
In S15, we accounted for this bias
as a systematic uncertainty of 5\% on the \mv\ lensing amplitude. 
In this analysis, we correct the cross talk between detectors at the time stream level as described
in H18.
With this correction, cross talk is suppressed by more than an order of magnitude. 
With the 5\% uncertainty from S15 being an upper limit before this suppression, we conclude that the residual
cross talk introduces negligible bias to our lensing amplitude measurement. 

Before binning into maps, 
we calibrate the time streams relative to each other using observations of the HII region
RCW38 and an internal chopped thermal source.
The per-detector polarization angles are calibrated based on measurements 
from a polarized thermal source~\citep{crites14}. 
After that, the time stream data are binned into $T/Q/U$ maps with $1^{\prime} \times 1^{\prime}$ pixels
using the oblique Lambert azimuthal equal-area projection. 

We apply the following corrections to the coadded map of the individual observations to
obtain the final map: $T \rightarrow P$ monopole subtraction, global polarization rotation, 
absolute calibration, and source and boundary masking. 
We measure $T \rightarrow P$ leakage by taking a weighted average over multipole space
 of the cross correlation of the temperature map with either the $Q$ or the $U$ map.
We find leakage factors of $\epsilon^Q=0.018$ and $\epsilon^U=0.008$. 
We obtain their uncertainties from the spread of leakage factors derived from
100 different half-splits of the data and find the fractional uncertainties to be 0.1\% and 0.3\% respectively. 
We deproject a monopole leakage term from the maps by subtracting a copy of the temperature 
map scaled by these factors from the $Q$ and $U$ maps.

Assuming \LCDM\ cosmology, we expect the cross spectra $TB$ and $EB$
to be consistent with zero. 
Therefore, we can estimate and apply the global polarization angle rotation needed to 
minimize the $TB$ and $EB$ correlations~\citep{keating13}.
We find the rotation angle to be  $0.63^{\circ} \pm 0.04^{\circ}$ and rotate the 
$Q$ and $U$ maps accordingly.

The final absolute calibration is obtained by comparing the final coadded \sptpol\ $T$ map with \planck\
over the angular multipole range $600 < \ell < 1000$.
Specifically, we take the ratio of the cross spectrum of two half-depth \sptpol\ maps to the cross spectrum
of full-depth \sptpol\ and \planck\ maps and require the ratio to be consistent
with 1 to determine the calibration factor.
We estimate the polarization efficiency (or polarization calibration factor) similarly, by 
comparing a full-depth \sptpol\ $E$-mode map with the \planck\ $E$-mode map. 
The temperature and polarization calibration factors (\tcal\ and \pcal) as derived
are $0.9088$ and $1.06$ respectively (see H18 for details).
We apply \tcal\ to the temperature map and $\tcal\times\pcal$ to the polarization maps
to obtain calibrated temperature and polarization maps. 
In the legacy \planck release, their polarization efficiency estimates are found to be potentially biased
at the $1-2\%$ level~\citep[see Table 9 of][]{planck18-3}. 
To circumvent potential biases when we tie our polarization map calibration to \planck's 
$E$-mode map, 
we adjust \pcal\ by a multiplicative factor obtained  from H18 without using \planck's polarization maps. 
In H18, \tcal\ and \pcal\ were free parameters in the fits of the
\texttt{PlanckTT+SPTpol EETE} dataset to the 
\LCDM + foregrounds + nuisance parameters model, 
with priors set by the \tcal\ and \pcal\ values and uncertainties
derived from the comparison against \planck\ described above. 
In this work, 
we multiply \pcal\ by the best-fit \pcal\ parameter (1.01) from H18 (see Table 5 of H18). 
In~\refssec{sys}, we use the posterior uncertainties of the \tcal\ and \pcal\ parameters
to  quantify their contributions to the systematic uncertainties of the lensing amplitude measurements. 

We apply a mask that defines the boundary of the $T/Q/U$ maps to downweight the noisy field edges. 
In addition, we mask bright point sources in the map. 
We use a $5'$ radius to mask point sources from the SPT-SZ catalog by
Everett et al. {\it in prep.}
with flux density above 6\,mJy at either 95\,GHz or 150\,GHz
that are in the 500\,deg$^2$ footprint.
We use a $10'$ radius for point sources with flux density
greater than 90\,mJy at either 95\,GHz or 150\,GHz. 
Clusters detected in~\cite{bleem15} within the 500\,deg$^2$ survey footprint 
are masked using a $10'$ radius.
The mask has a top hat profile for both the boundary and the sources, 
and mode mixing due to the mask is suppressed by the inverse-variance map-filtering step (\refssec{cmbmapfilt}).  
We compare this lensing measurement with one using a cosine-tapered apodization on the mask edges 
in~\refssec{sys} and find them to be consistent.  

The noise levels in the coadded maps are 11.8\,$\mu$K--arcmin in temperature and 
8.3\,$\mu$K--arcmin in polarization over the multipole range
$1000 < \ell < 3000$\footnote{The temperature map noise is higher than the polarization map
noise in this multipole range because of atmospheric noise contributions.}.
The map depth is similar to that of the 100\,deg$^2$ field used in 
S15\footnote{The noise levels provided in S15 were estimated without the \tcal\
and \pcal\ corrections and are thus higher.}. 
Since this map covers five times the sky area, the sample variance of the lensing 
power spectrum is reduced and hence the lensing spectrum measurement precision
is improved. 

\subsection{Simulations}
\label{ssec:sims}
We use simulated skies to estimate the mean-field bias $\bar{\phi}^{\rm MF}$ 
to the lensing potential map (\refssec{lenspo}), 
to correct the analytical response of the lensing estimator 
(\refssec{lenspo}), to correct the
expected biases (\rdn and \none, see~\refssec{lensspec}) in the raw lensing power spectrum,
and to estimate the uncertainties of the lensing measurements (\refsec{results}). 

The simulated skies contain the CMB, foregrounds, and instrumental noise. 
The input cosmology for generating the CMB is the best-fit $\Lambda$CDM model 
to the \planck~$\texttt{plikHM\_TT\_lowTEB\_lensing}$ dataset~\citep[second column in Table 4 of][]{planck15-13}.
Using the best-fit cosmology, we use CAMB\footnote{http://camb.info} to generate
theory spectra,
which are then fed into {\sc Healpix}~\citep{gorski05} to generate
the spherical harmonic coefficients $a_{\ell m}$ for $T, E, B$, and the lensing potential $\phi$.
The CMB $a_{\ell m}$ are projected to maps and lensed by $\phi$ using {\sc LensPix}~\citep{lewis05}. 
The lensed maps are then converted back to $a_{\ell m}$, at which point 
foreground fluctuations are added, and the $a_{\ell m}$ are multiplied by the $\ell$-space 
instrument beam window function. 
The $a_{\ell m}$ are subsequently projected on 
an equidistant cylindrical projection (ECP) grid for mock observation, 
which produces mock skies that are processed identically as the real sky. 

The foreground components are generated as Gaussian realizations of 
model angular power spectra. 
We include power from thermal and kinematic Sunyaev-Zel'dovich effects (tSZ and kSZ), 
the cosmic infrared background (CIB), radio sources, and galactic dust. 
The amplitudes for tSZ, kSZ, and CIB components are taken from \cite{george15}, 
which has the same source masking threshold as this work. 
We use an amplitude $D_{\ell=3000}^{\rm tSZ+kSZ}$ = 5.66\,\muKsq\ with 
a model shape from \cite{shaw10} for the sum of tSZ and kSZ 
components\footnote{Given that we mask all clusters in the \citet{bleem15} 
catalog, this tSZ level is high by $\sim$2\,\muKsq. As a result, 
we non-optimally downweight high-$\ell$ modes in the temperature data map
and slightly over-estimate the $N_L^{(0)}$ noise (\refssec{lensspec}). 
We compare the analytic $N_L^{(0)}$ between $D_{\ell=3000}^{\rm tSZ+kSZ} = 5.66$\,\muKsq\
vs $D_{\ell=3000}^{\rm tSZ+kSZ} = 3.61$\,\muKsq\ (keeping all other inputs equal) 
and find that the latter is $\sim$2\% lower. 
Since there is both signal and noise variance in the uncertainty of the lensing amplitude
measurement, and the \tt\ lensing amplitude uncertainty is \ttPercent\ (see~\refsec{results}), 
the \tt\ lensing amplitude would at most be reduced by $\sim$2\% of \ttPercent,
which is negligible.}. 
We use $D_{\ell=3000}^{\rm radio} = 1.06$\,\muKsq\ with $D_{\ell}\propto \ell^2$ 
for the radio source component. 
We use  $D_{\ell=3000}^{{\rm CIB},P} = 9.16$\,\muKsq\ with $D_{\ell}\propto \ell^2$ 
for the unclustered CIB component and $D_{\ell=3000}^{{\rm CIB},cl} = 3.46$\,\muKsq\ 
with $D_{\ell}\propto \ell^{0.8}$ for the clustered CIB. 
By modeling these terms as Gaussian, we have neglected
the potential bias their non-Gaussianities can introduce to the lensing spectrum.
We treat this potential bias explicitly later in the analysis (see~\refssec{lensspec}).
We assume a 2$\%$ polarization fraction for all the Poisson-distributed (unclustered) 
components to model extragalactic polarized emission~\citep{seiffert07}. 
We model galactic dust power in temperature and polarization as a power law with
$D_{\ell}\propto \ell^{-0.42}$, with 
$D_{\ell=80}^{{\rm TT},dust} = 1.15$\,\muKsq,
$D_{\ell=80}^{{\rm EE},dust} = 0.0236$\,\muKsq, and 
$D_{\ell=80}^{{\rm BB},dust} = 0.0118$\,\muKsq~\citep{keisler15}. 

The instrumental noise realizations 
are generated by subtracting half of all the observations from the other half. 
By randomly grouping the observations into one of the halves, 
we create 500 different realizations of noise from the data themselves.
The noise realizations are added to the simulated skies after mock observation. 

We generate 500 lensed skies including CMB, foregrounds, and noise. 
All 500 skies are used for estimating the lensing spectrum bias \rdn.
We use 100 skies to estimate the mean-field ($\bar{\phi}^{\rm MF}$, see~\refssec{lenspo}): 
50 for each of the two lensing potential 
$\phi$ estimates that enter the lensing spectrum calculation 
($\hat{\phi}^{UV}_{\bL}$ and $\hat{\phi}^{XY}_{\bL}$ in ~\refeq{lensspec_cross}).
We use the other 400 to correct the analytical response of the lensing estimator 
and to obtain the statistical uncertainty of the lensing spectrum. 
In addition, we generate 500 unlensed skies with all other inputs being identical.
We use this set of simulations for testing the lensing pipeline and assessing
the probability of detecting lensing from unlensed skies.

The \none\ bias is estimated using a different set of 200 lensed skies. 
We start with 200 realizations of unlensed skies and divide them
into two groups.
We then lens one sky from each group by the same $\phi$ (\refssec{lensspec}). 
We do not add instrumental noise nor foregrounds to this set of simulations, 
because \none\ comes from correlations between the CMB and the lensing modes.
At the map-filtering step, we filter this set of simulations 
assuming the same level of foregrounds and noise as the other set of lensed skies.

\section{Lensing Analysis}
\label{sec:lensing}
In this section, we describe the lensing pipeline that produces an unbiased estimate 
of the lensing potential and of the lensing power spectrum.
This analysis follows that of S15 except for the treatment of the MC bias
and foreground subtraction in the lensing spectrum. 
We summarize the steps here and refer the reader to S15 for details. 

\subsection{Estimating the lensing potential}
\label{ssec:lenspo}
The unlensed CMB sky is well approximated by a statistically isotropic, Gaussian 
random field. 
Gravitational lensing breaks the statistical isotropy of the fluctuations and 
introduces correlations between the otherwise uncorrelated Fourier
modes of the CMB temperature and polarization maps. 
We use these correlations to estimate the underlying lensing 
potential with pairs of filtered maps using the quadratic estimator of~\citet{hu02a}: 
\beq
\label{eq:phi_bar}
\bar{\phi}^{XY}_{\bL} = \int{d^2\bl W^{XY}_{\bl,\bl-\bL}}
\bar{X}_{\bl}\, \bar{Y}^{*}_{\bl-\bL},
\eeq
where $\bar{X}$ and $\bar{Y}$ are filtered $T$, $E$ or $B$ fields as outlined in~\refssec{cmbmapfilt}, 
$W$ is a weight function unique to each input pair of $XY$ maps, 
$\bl$ are modes of the CMB, and $\bL$ are 
modes of the lensing potential.
We form lensing potentials from $XY \in [TT, TE, EE, EB, TB]$
and modify
$W$ from \cite{hu02a} by replacing the unlensed CMB spectra with 
lensed spectra to reduce a higher-order bias~\citep[$N_L^{(2)}$,][]{hanson11}. 
As written, $\bar{\phi}$ is a biased estimate of $\phi$. 
To arrive at an unbiased estimate of $\phi$, we remove an additive
bias (the mean-field, MF) and normalize
the estimator by the response (defined below). 

The mean-field can arise from masking and inhomogeneous noise -- sources that introduce
mode coupling across angular scales.
These mode couplings persist even when the CMB and $\phi$ realizations are different.
We can therefore estimate the mean-field by averaging $\bar{\phi}$ from many realizations
of input lensed CMB maps. 
For the $L$ range considered in this work, the mean field is subdominant compared
with the lensing signal spectrum. 
Specifically, it is $\lesssim$ 30\% of the lensing power spectrum in the first $L$ bin. 
It grows larger towards larger angular scales, dominated by the effect of the mask.

We construct the response using an analytical calculation corrected 
by simulations. 
The analytic response of the estimator $R^{XY,{\rm Analytic}}_{|\bL|} $ is
\beq
\label{eq:phi_resp}
R^{XY,{\rm Analytic}}_{|\bL|} = \int{ d^2\bl\, W^{XY}_{\bl,\bl-\bL}\times W^{XY}_{\bl,\bl-\bL} \mathcal{F}^{X}_{\ell} \mathcal{F}^{Y}_{\bl-\bL} } \,,
\eeq
where $\mathcal{F}^{X}_\ell = (C^{XX}_\ell + N^{XX}_\ell)^{-1} $
describes the diagonal approximation to the filter applied to the input CMB maps
(\refssec{cmbmapfilt}). 
Because this filter assumes spatial stationarity of the statistics of the signal and the noise, it does not
account for nonstationary effects from e.g. the boundary and source mask and 
causes the response to be slightly misestimated. 
To account for this, we use simulations to estimate the correction to 
the analytic response.
The total response is thus
$R^{XY}_{|\bL|} =R^{XY,{\rm Analytic}}_{|\bL|} \times R^{XY,{\rm MC}}_L$, 
with $R^{XY,{\rm MC}}_L$ denoting the correction estimated from simulations.
We extract $R^{XY,{\rm MC}}_L$ by first taking the cross spectrum between 
the input $\phi^{\rm in}$ and the intermediate $\phi$ estimate $\hat{\phi}^{\prime XY}$, 
which has been mean-field subtracted and normalized by the analytic response: 
$\hat{\phi}^{\prime XY}_{\bL} = (\bar{\phi}^{XY}_{\bL} - \bar{\phi}^{XY,{\rm MF}}_{\bL})/R^{XY,{\rm Analytic}}_{|\bL|}$.
We then take the ratio of the average of the cross spectra 
$\langle \phi^{\rm in \ast}_{\bL}\hat{\phi}^{\prime XY}_{\bL}  \rangle$
averaged within each L annulus
with the input 
spectrum over 400 simulation realizations.
Similar to S15, we find the  $R^{XY,{\rm MC}}_L$ to be a $\leqslant 10\%$ correction.

The normalized and mean-field-corrected $\hat{\phi}$ is 
\beq
\label{eq:phi_hat}
\hat{\phi}^{XY}_{\bL} = \frac{1}{R^{XY}_{|\bL|}}
  (\bar{\phi}^{XY}_{\bL} - \bar{\phi}^{XY,{\rm MF}}_{\bL}) \,
\eeq
for the individual estimators where $XY \in [TT, TE, EE, EB, TB]$. 
In this work, we use \tt\ to denote the lensing potential and spectrum constructed
using only temperature data, i.e. $\hat{\phi}^{TT}_{\bL}$ and its spectrum. 
We use \pp\ to denote the potential and spectrum constructed using only polarization data,
and \mv\ to denote the potential and spectrum constructed using both temperature
and polarization data. 

To construct the combined minimum-variance (\mv) and polarization-only (\pp) estimators,
we first form a weighted average of the input estimators to form the intermediate 
$\hat{\phi}^{\prime}_{\bL}$. 
We use the inverse-noise variance of the input estimators as weights, which
are approximated by their analytical responses.
We then extract $R^{\rm MC}_L$ by forming cross spectra between 
$\hat{\phi}^{\prime\rm \mv}_{\bL}$
or $\hat{\phi}^{\prime\rm \pp}_{\bL}$ with the input $\phi^{\rm in}_{\bL}$.
Putting it all together, the unbiased \mv\ and \pp\ lensing potentials are constructed as 
\beq
\label{eq:phi_mv}
\hat{\phi}_{\bL} = 
 \frac{1}{R^{MC}_L } \frac{ \sum_{XY}  \bar{\phi}^{XY}_{\bL} -  \bar{\phi}^{XY, {\rm MF}}_{\bL} }{ \sum_{XY} R^{XY,{\rm Analytic}}_{|\bL|} } \,,
\eeq
where $XY \in [TT, TE, EE, EB, TB]$ for the \mv\ estimator and $XY \in [EE, EB]$ for 
the \pp\ estimator.

\subsection{Input CMB map filtering}
\label{ssec:cmbmapfilt}
We filter our input maps with an inverse-variance (C$^{-1}$) filter, 
which is derived such that the variance of a lensing field
reconstructed by a quadratic estimator is minimized.
The filter is constructed identically as in S15: 
we assume the data maps to be composed of three components:
sky signal, ``sky noise", and pixel domain noise.
The sky signal and noise are modeled in the Fourier domain and they include
the CMB, astrophysical foregrounds, and atmospheric noise. 
The pixel domain noise is modeled as white, uncorrelated, and spatially 
nonvarying inside the mask. 
Concretely, we solve for the inverse-variance filtered Fourier modes $\bar{X}$ in the
following expression using conjugate-gradient-descent: 
\beq
\left[ S^{-1} + P^{\dagger} n^{-1} P \right] S \bar{X} = P^{\dagger} n^{-1} d \,.
\eeq
Here $S \equiv C^{XX}_{\ell} + N^{XX}_{\ell}$ describes the sky signal and noise
components, $n^{-1}$ is the inverse of the map noise variance and is zero for
masked pixels, $P$ applies the filter transfer function and inverse-Fourier 
transforms the map from pixel space to Fourier space. 
$P$ additionlly transforms $Q/U$ to $E/B$ for polarization maps, 
and $d$ denotes the input pixel-space $T/Q/U$ maps. 

This filter approaches the simple form 
$\bar{X} = S^{-1} P^{\dagger} d = (C^{XX}_\ell + N^{XX}_\ell)^{-1} X $
in regions far away from the mask boundary. 
To see that, one can absorb the pixel-domain noise into the $N^{XX}_\ell$
term (which is valid under our model of the data for pixels
far from the mask boundary); in the limit where $n \rightarrow 0$, the 
diagonal form is exact. 

The inputs to this filter are as follows. 
For the sky signal component of $S$, we use the same lensed
CMB and astrophysical foreground spectra used to
generate simulations (\refssec{sims}).
For the sky noise component, we take the averaged power spectra from 
the temperature and polarization noise realizations and from them 
subtract noise floors of 7\,$\mu$K--arcmin to form $N^{XX}_\ell$.
We set 7\,$\mu$K--arcmin as the white noise level
of the pixel-domain noise of the $T$, $Q$, and $U$ maps.  
The filter transfer function $P$ includes time stream filtering, beam, and pixel-window function.
We approximate the time stream filtering transfer function using simplified simulations
that capture the lost modes along the $\ell_x = 0$ axis due to our scan strategy. 
We measure the SPT beam using the response of the detectors to Venus, 
as was done in H18. 
In addition, we fit the measured beam with a function of form
$B(\ell) = \sum_{i=1,2} \,A_i {\rm exp}(- 0.5 (\ell/\ell_i)^{p_i})$ to obtain a smooth profile at $\ell < 400$. 
We incorporate the pixel-window function for the map pixel size of $1^{\prime} \times 1^{\prime}$.
Our input CMB map $\bl$ range for this analysis is $ | \bl_x |  > 100$ set by the time stream 
$\ell$-space high-pass filter and $| \bl | < 3000$ set by concerns of foreground
contamination (\refssec{sys}).

\subsection{Estimating the lensing potential power spectrum}
\label{ssec:lensspec}
Upon obtaining the unbiased estimate of the lensing potential $\hat{\phi}$, we calculate
the raw power spectrum of the lensing potential  $C^{\hat{\phi}^{UV} \hat{\phi}^{XY}}_{L}$ 
by forming cross spectra of $\hat{\phi}^{XY}_{\bL}$
and $\hat{\phi}^{UV}_{\bL}$ where $UV, XY \in [TT, TE, EE, EB, TB]$:
\beq
\label{eq:lensspec_cross}
 C^{\hat{\phi}^{UV} \hat{\phi}^{XY}}_{L}
 \equiv 
 f_{\rm mask}^{-1}\sum_{|\bL| = L}\langle \hat{\phi}^{UV}_{\bL} \,\, \hat{\phi}^{*}\,^{XY}_{\bL} \rangle \, ,
\eeq
where $f_{\rm mask}$ is the average value of the fourth power of the mask. 

\begin{figure}
\begin{center}
\includegraphics[width=0.48\textwidth]{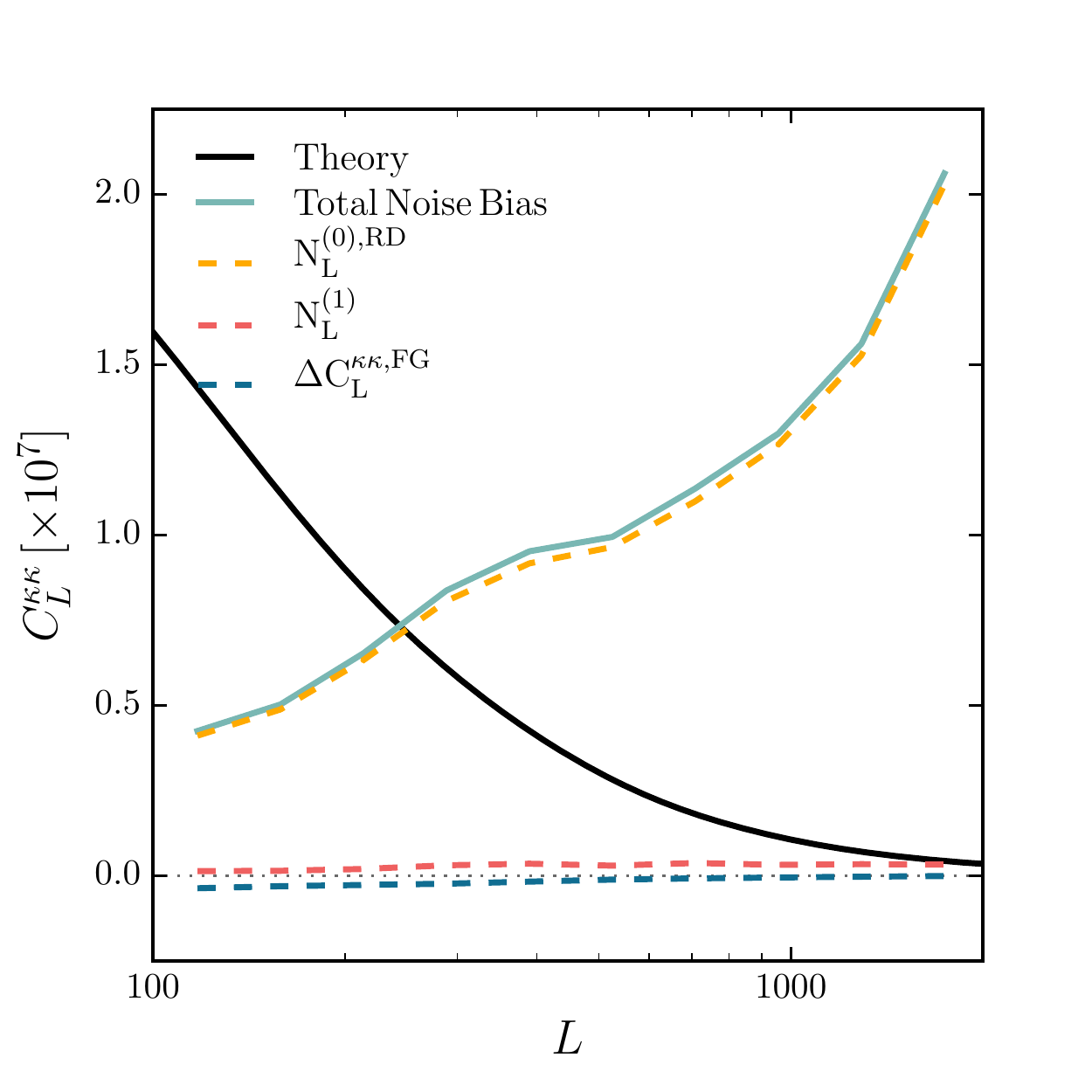}
\caption{Bias terms subtracted from the raw \mv\ lensing spectrum: \rdn, \none 
(converted to $C_L^{\kappa\kappa}$ units, ~\refeq{convergence}), 
and foreground bias $\Delta C_{L}^{\kappa \kappa, {\rm FG}}$. 
The theoretical lensing convergence spectrum for the fiducial cosmology is shown in black.
The disconnected term in the CMB 4-point function \rdn\ is shown in dashed yellow (\refeq{rdn0}).
The spurious correlated power between the CMB and the lensing potential \none\ is shown
in dashed pink (\refeq{n1}).
The sum of these two terms is labeled ``Total Noise Bias" in the figure in solid green. 
The foreground bias from the tSZ trispectrum, CIB trispectrum, and the tSZ and CIB correlation
with $\kappa$ is shown in dashed dark blue (\refssec{lensspec}).
The total noise bias is also an estimate of the noise in the reconstructed $\phi$ map.
Lensing modes with values of $L$ for which the total bias is less than the signal spectrum
are measured at signal-to-noise greater than unity. 
For this measurement, that includes all modes with $L \lesssim$ 250,
or angular scales of roughly a degree and larger. 
}
\label{fig:bias_amp}
\end{center}
\end{figure}

This power spectrum is biased, and 
we correct four sources of bias in this analysis.
Two of them, the \nzero\ and \none\ biases, arise from spurious correlations
of the input fields.
The third source of bias comes from 
foregrounds, $\Delta C_{L}^{\phi \phi, {\rm FG}}$.
Finally, we correct for a multiplicative bias due to 
higher-order coupling of the source mask $f_{\rm PS}$. 
Therefore, the unbiased lensing spectrum is estimated as
\begin{multline}
\label{eq:lensspec}
\hat{C}^{\phi \phi}_{L} =  f_{\rm PS} \left[ C^{\hat{\phi} \hat{\phi}}_{L} - \rdn - \none \right] -  \Delta C_{L}^{\phi \phi, {\rm FG}} \,.
\end{multline}
\reffig{bias_amp} shows \rdn, \none, and  $\Delta C_{L}^{\phi \phi, {\rm FG}}$
for the \mv\ reconstruction. 
\rdn\ and \none\ are similar to those in S15 because the CMB map noise levels 
for these two works are about the same. 
We describe how each bias term is estimated in the following paragraphs. 

The lensing spectrum is a 4-point function, or trispectrum, of the observed fields and thus contains, 
in addition to the connected term caused by lensing, 
a disconnected term from correlations of Gaussian fields (\nzero).
Secondary contractions of the trispectrum give rise to 
connected terms that also bias the lensing spectrum (\none). 
It is called \none\ because it is first order in $C_L^{\phi\phi}$~\citep{kesden03, hanson11}.
We estimate the power of the \nzero\ and \none\ terms using simulations
and subtract them from the raw lensing spectrum.

We estimate the disconnected term using a realization-dependent method (\rdn, \citealt{namikawa13}).
The standard \nzero\ is estimated by computing the lensing spectra with
two input maps that form $\hat{\phi}$ from different simulation realizations.
Since input maps from different realizations do not share the lensing potential, 
the resultant power in the lensing spectra comes from spurious correlations
of the maps. 
With the realization-dependent method, we use a combination of simulation
and data maps.
Specifically, on top of the mix of simulation realizations, 
we also construct lensing potentials using data in 
one of the input maps. 
Accounting for all the different combinations, we have

\begin{multline}
\rdn = \\
\begin{aligned}
\Big \langle
  & +\cmbLppHat[\bar{U}_{\rm d},  \bar{V}_{\rm MC},  \bar{X}_{\rm d},   \bar{Y}_{\rm MC}]
    +\cmbLppHat[\bar{U}_{\rm MC}, \bar{V}_{\rm d},   \bar{X}_{\rm d},   \bar{Y}_{\rm MC}]  \\
  & +\cmbLppHat[\bar{U}_{\rm d},  \bar{V}_{\rm MC},  \bar{X}_{\rm MC},  \bar{Y}_{\rm d} ]
    +\cmbLppHat[\bar{U}_{\rm MC}, \bar{V}_{\rm d},   \bar{X}_{\rm MC},  \bar{Y}_{\rm d} ]  \\
  & -\cmbLppHat[\bar{U}_{\rm MC}, \bar{V}_{\rm MC'}, \bar{X}_{\rm MC},  \bar{Y}_{\rm MC'}] \\
  & -\cmbLppHat[\bar{U}_{\rm MC}, \bar{V}_{\rm MC'}, \bar{X}_{\rm MC'}, \bar{Y}_{\rm MC}] 
  \Big \rangle_{\rm MC,MC'} \,,
\end{aligned}
\label{eq:rdn0}
\end{multline}
where $C^{\hat{\phi}^{UV} \hat{\phi}^{XY}}_{L}$ is expressed as 
$\cmbLppHat[\bar{U},\bar{V},\bar{X},\bar{Y}]$ to more clearly indicate the sources of the input maps. 
Here $MC$ and $MC'$ denote simulation skies from different realizations, and $d$
denotes the data map. 
The realization-dependent method produces a better estimate of the disconnected term
because it reduces the bias from
the mismatch between the fiducial cosmology used for generating the simulations 
and that in the data. 
In addition, this method reduces the covariance between the lensing potential
bandpowers.

The second bias term \none\ comes from spurious correlations between the CMB and the 
lensing potential and is proportional to $C_L^{\phi\phi}$. 
We estimate it using pairs of simulated skies that have different realizations of 
unlensed CMB lensed with the same $\phi$: 
\begin{multline}
\label{eq:n1}
\none = \\
\begin{aligned}
\Big \langle
  & +\cmbLppHat[\bar{U}_{\phi^1,{\rm MC}},\bar{V}_{\phi^1,{\rm MC'}},\bar{X}_{\phi^1,{\rm MC}},\bar{Y}_{\phi^1,{\rm MC'}}] \\
  & +\cmbLppHat[\bar{U}_{\phi^1,{\rm MC}},\bar{V}_{\phi^1,{\rm MC'}},\bar{X}_{\phi^1,{\rm MC'}},\bar{Y}_{\phi^1,{\rm MC}}] \\
  & -\cmbLppHat[\bar{U}_{\rm MC},\bar{V}_{\rm MC'},\bar{X}_{\rm MC},\bar{Y}_{\rm MC'}] \\
  & -\cmbLppHat[\bar{U}_{\rm MC},\bar{V}_{\rm MC'},\bar{X}_{\rm MC'},\bar{Y}_{\rm MC}] 
  \Big \rangle_{{\rm MC,MC'}}\,,
\end{aligned}
\end{multline}
where the subscript $\phi^1$ denotes CMB maps lensed by the same $\phi$ realization. 

We account for biases to the lensing spectrum due to foregrounds in the temperature map.
As studied in~\cite{vanengelen14a}, both tSZ and CIB have a trispectrum, which leads to 
a response in the lensing power spectrum. 
This bias enters the lensing spectrum through the 4-point function of the temperature map,
thus modifying the \tt\ spectrum and the \mv\ spectrum.  
Additionally, since both tSZ and CIB trace the same large-scale structure as the lensing
field, 
the non-Gaussianities of both fields can mimic lensing and couple through
the $\phi$ estimator in a coherent way that correlates with $\phi$,
forming a nonzero bispectrum of the matter density field (denoted by tSZ$^2$-$\kappa$
and CIB$^2$-$\kappa$). 
This effect biases the \mv\ lensing spectrum and spectra from 
pairs of estimators of the form $\left < \phi^{TT} \phi^{UV}\right>$,
where $UV \in [TT, TE, EE, EB, TB]$.
The level of foreground bias is scale dependent.
It is negative and close to flat at $\sim$2\% for $L < 1300$ for 
the total bias contributed from tSZ and CIB through their trispectra and correlations with $\phi$.
As $L$ increases, the magnitude of this bias decreases and reaches a null at about $L$ of 2000. 
We subtract the relevant terms 
from the \tt\ and \mv\ spectra by using the bias estimates from \cite{vanengelen14a}.
Specifically, we subtract the total foreground bias coming from the tSZ trispectra, CIB trispectra, 
tSZ$^2$-$\kappa$, and CIB$^2$-$\kappa$ from the \tt\ lensing spectrum. 
For the \mv\ spectrum, we subtract the total foreground bias that enters through
$\left < \phi^{TT} \phi^{TT}\right>$, and tSZ$^2$-$\kappa$ and CIB$^2$-$\kappa$ biases
that enter through  $\left < \phi^{TT} \phi^{UV}\right>$ with $UV \in [TE, EE, EB, TB]$.
We compute the bias fraction by forming a weighted average of the foreground biases to
$\left < \phi^{TT} \phi^{TT}\right>$
and $\left < \phi^{TT} \phi^{UV}\right>$. 
We use as weights the fractional contribution of the estimators to the \mv\ estimator.
The size of this bias in the \mv\ spectrum is shown in~\reffig{bias_amp}.

MC bias describes the difference between the recovered amplitude in simulations 
and the input spectrum. 
It is typically found to be small~\citep[e.g.][]{sherwin16, planck18-8}.
In S15, we found that the mean \mv\ amplitude $A_{\rm \mv}$ from simulations was 3\% below
unity and treated this discrepancy as a source of systematic uncertainty. 
In this work, we find that the main source of our MC bias comes from the inclusion of 
the point source and cluster mask.
When analyzing the set of simulations with the source mask removed (while keeping the 
boundary mask), we are able to recover $A_{\rm \mv} = 1$ to within 1\,$\sigma$ of 
the standard error of 400 sky realizations ($0.058/\sqrt{400} = 0.0029$) for the multipole
range we report in this work.
Furthermore, comparing the mean recovered spectrum of the \mv, \pp, and \tt\ estimators with
and without the source mask applied to the input maps, we observe a relatively constant multiplicative
offset in the range of $100 < L \lesssim 800$. 
We therefore conclude that the main source of our MC bias is due to higher-order coupling 
generated from the presence of source masks in the map that is not accounted for
by $f_{\rm mask}$. 
We construct a multiplicative correction for this MC bias $f_{\rm  PS}$ as an inverse of the mean of
the simulation lensing amplitudes $A_{\rm lens}$  estimated from $100 < L < 602$: 
\beq
\label{eq:fmask_pt}
f^{XY, UV}_{\rm PS} = {\langle A^{XY, UV}_{\rm lens} \rangle }^{-1}\,.
\eeq
To check the stability of this estimate, we vary the maximum $L$ range used between
$L$ of 446 and 813 and find 
the resultant simulation $A_{\rm lens}$ 
from all pairs of estimators to be consistent with unity to within 2\,$\sigma$ of 
their standard errors. 
The correction from this step is about 5\%.
Specifically, $f_{\rm PS}$ for the \mv, \pp, and \tt\ are
$f^{\rm\mv}_{\rm PS}$ = 1.05,  $f^{\rm \pp}_{\rm PS}$ = 1.05, and 
$f^{\rm\tt}_{\rm PS}$ = 1.07. 
For $L < 100$, the mean of the simulation spectra with the source mask removed is 
$5-10\%$ below the input spectrum.
Therefore, this MC bias correction is not applicable for multipoles below 100, 
and we do not report results below $L $ of 100. 

\begin{figure*}
\begin{center}
\begin{minipage}{1\linewidth}
\begin{subfigure}{}
\includegraphics[width=0.98\textwidth]{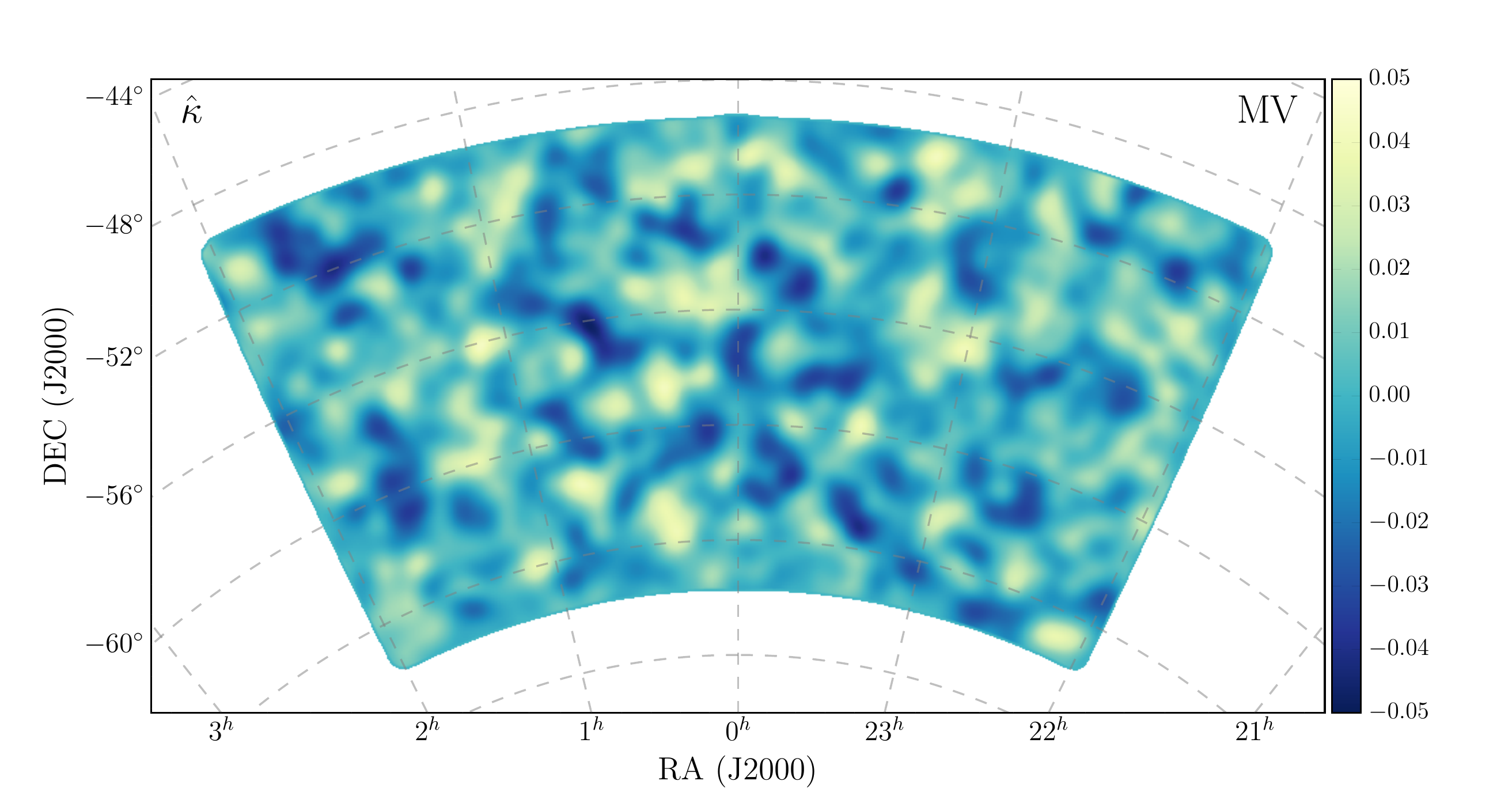}
\end{subfigure}
\end{minipage}

\begin{minipage}{1\linewidth}
\begin{subfigure}{}
\includegraphics[width=0.475\textwidth]{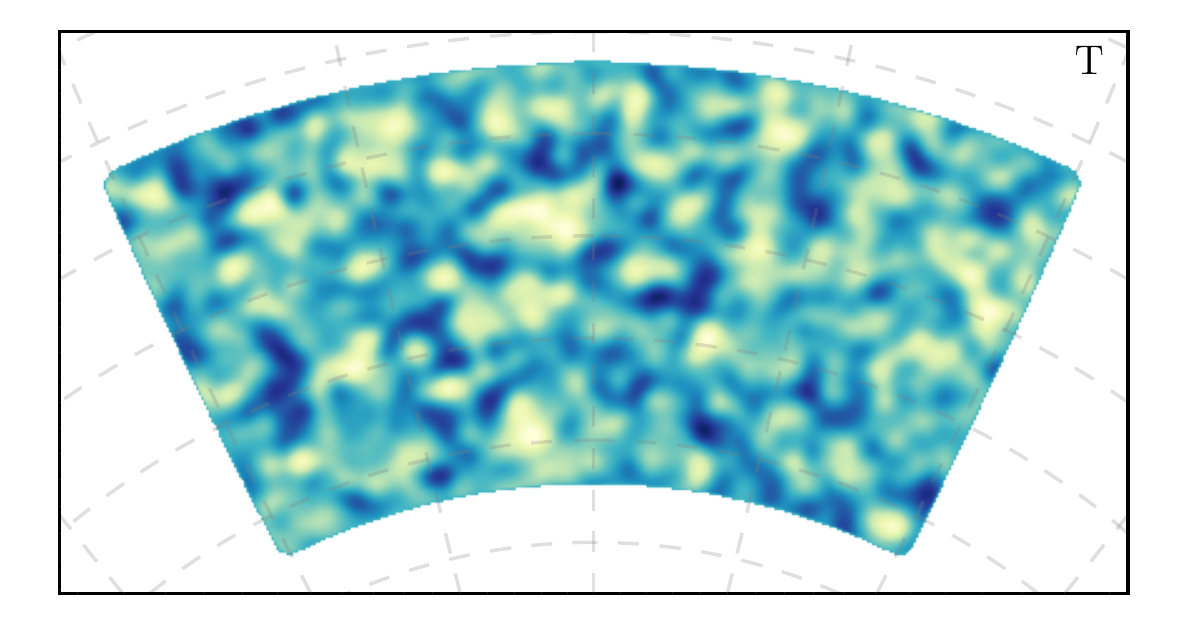}
\includegraphics[width=0.475\textwidth]{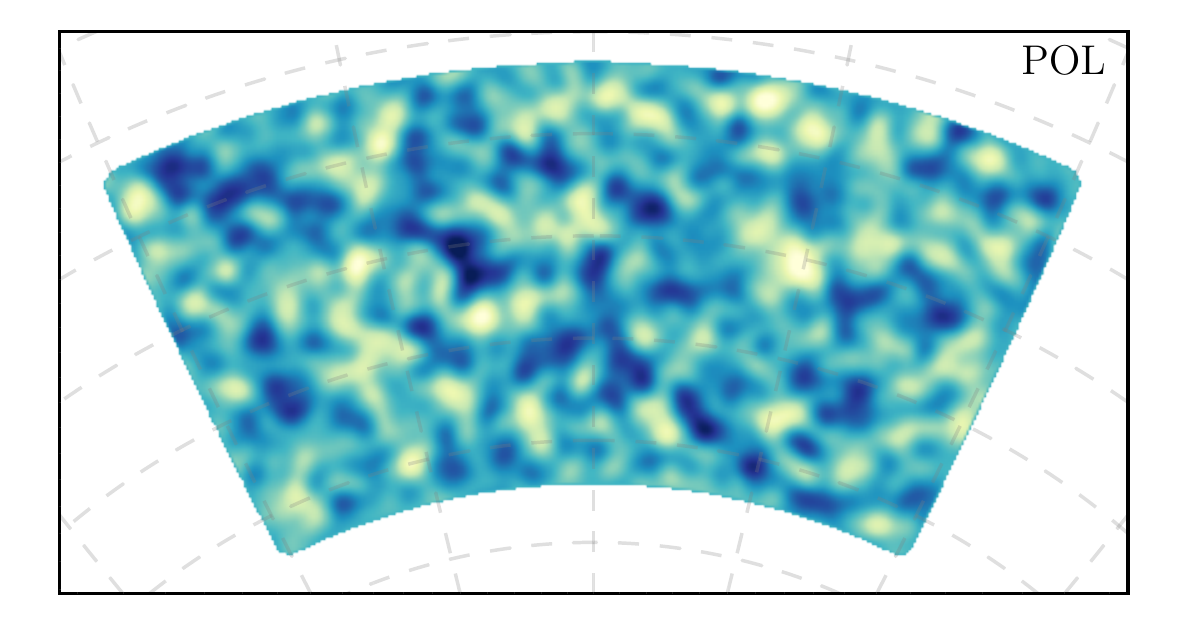} 
\end{subfigure}
\end{minipage}

\caption{Lensing $\kappa$ map reconstructed from the \sptpol 500\,deg$^2$ field data, smoothed by a 1-degree FWHM Gaussian to highlight the signal-dominated modes. 
{\bf Top:} The $\kappa$ map from the \mv\ lensing estimator, which combines all temperature and polarization information.
{\bf Bottom:} $\kappa$ maps from the temperature estimator (\tt, left) and from combining the polarization estimators $EE$ and $EB$ (\pp, right).
At the angular scales shown, the \pp\ estimator recovers the lensing potential with slightly higher S/N than the \tt\ estimator. 
Therefore, it has higher weight in the \mv\ combination and traces the fluctuations of the \mv\ $\kappa$ map with higher fidelity. 
}
\label{fig:kappa_mv}
\end{center}
\end{figure*}

In constructing our simulations and applying the quadratic estimator to recover
the lensing potential, we have assumed $\phi$ to be Gaussian.
However, nonlinear structure growth and post-Born lensing would introduce
non-Gaussianities to $\phi$~\citep{boehm16, pratten16}. 
These non-Gaussianities produce the so-called $N_L^{(3/2)}$ bias, as studied in~\cite{boehm16, boehm18, beck18}.
For the $L$ range considered, the size of this bias is $\sim$0.5\% 
for the temperature reconstruction and is negligible for the $EB$ reconstruction
for input CMB maps similar in noise levels and multipole range to 
those in this work~\citep{boehm18, beck18}. 
These are small compared with our other sources of uncertainties, and
we neglect them in our results. 

We report our results in binned bandpowers. 
First, we derive the per-bin amplitude as
the ratio of the unbiased lensing spectrum to the input theory spectrum:
\beq
\label{eq:amp_def}
A_b^{UV XY} \equiv \frac{C^{\phi^{UV} \phi^{XY}}_b }
              {C^{\phi \phi, {\rm theory}}_b } \,,
\eeq
with the subscript $b$ denoting a binned quantity;
$C_b$ is a weighted average of the $C_L$ inputs for $L$ inside the boundaries of the bin, with the weights $w$
designed to maximize signal-to-noise: 
$w^{UV, XY}_L = C^{\phi \phi, {\rm theory}}_L / {\rm Var}( C^{\hat{\phi}^{UV} \hat{\phi}^{XY}}_{L} )$.
We obtain the variance from the corresponding set of simulation cross spectra.

We report the bandpowers in lensing convergence ($\kappa$) instead of the
lensing potential $\phi$.
The convergence field is $-1/2$ of the divergence of the deflection field, which is the gradient
of the lensing potential $\phi$~\citep{lewis06}:
\beq
\label{eq:convergence}
\kappa = -\frac{1}{2} \nabla^2 \phi\, .
\eeq
In Fourier space, they are related by $\kappa_L = (L (L+1)) \phi_L /2$.
The reported bandpowers are derived as the product of the data amplitude $A_b$
and the input theory spectrum at bin center $L_b$,
\beq
\label{eq:bandpowers}
\hat{C}^{\kappa\kappa}_{L_b} \equiv  \frac{( L_b (L_b + 1))^2}{4} A_b \, C^{\phi \phi, {\rm theory}}_{L_b} \,.
\eeq

The overall lensing amplitude for each estimator is calculated identically as 
the per-bin amplitude in~\refeq{amp_def} using the whole reported $L$ range. 

\begin{figure*}
\begin{center}
\includegraphics[width=0.35\textwidth]{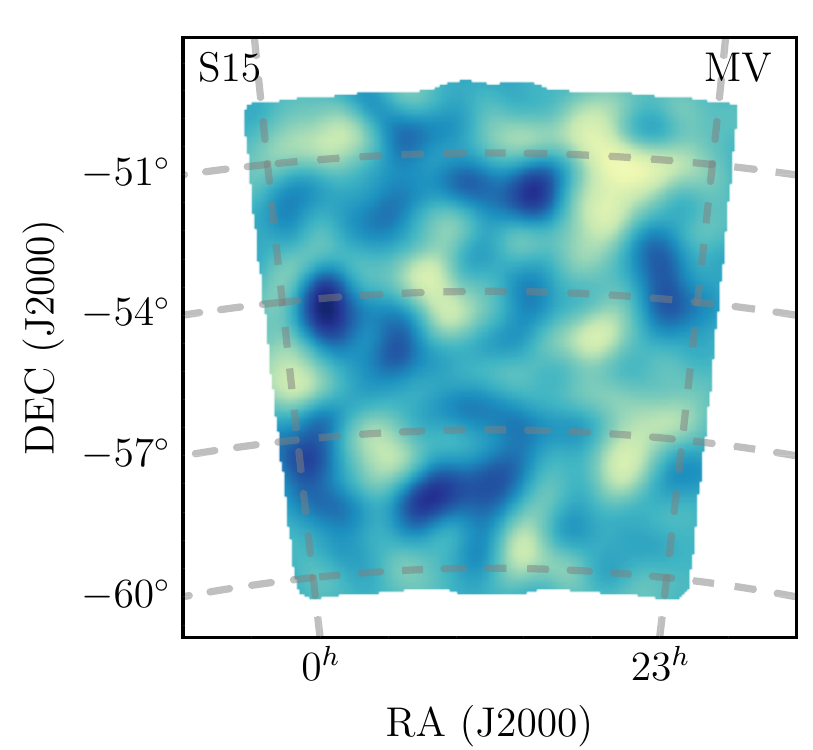} 
\includegraphics[width=0.35\textwidth]{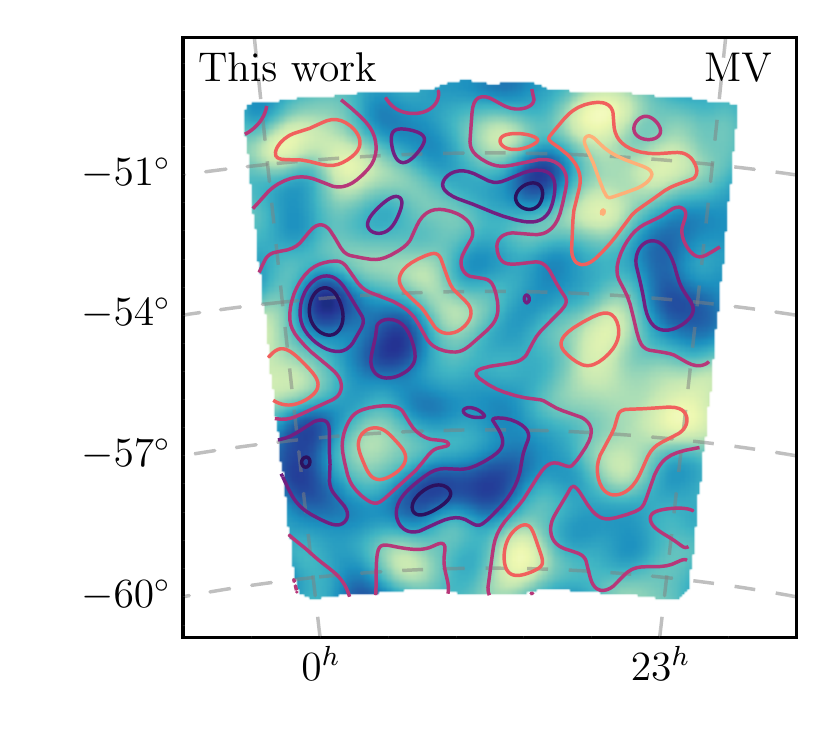}
\includegraphics[height=0.325\textwidth]{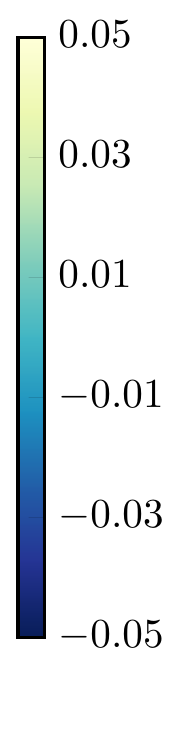}
\caption{
Comparison with the S15 $\kappa$ map:
{\bf Left:} The $\kappa$ map reconstructed by S15 using the \mv\ estimator with maps from
\sptpol 100\,deg$^2$ field observations (the sign is opposite from that shown in S15 because of 
a sign error in plotting in S15).
{\bf Right:} Cutout of the 100\,deg$^2$ field from the \mv\ $\kappa$ map in this analysis,
overlaid with contours from the S15 \mv\ $\kappa$ at $[-0.048, -0.032, -0.016, 0, 0.016, 0.032, 0.048]$.
Both maps are smoothed by a 1-degree FWHM Gaussian. 
Data that enter the S15 reconstruction were taken before the 500\,deg$^2$ survey, 
and thus are independent from the data used in this analysis.
Since modes shown are measured to S/N greater than unity, 
the fluctuations of the two maps visually tracing each other serves as a consistency
check.}
\label{fig:kappa_s15}
\end{center}
\end{figure*}

Differences between the fiducial cosmology and the cosmology of the \sptpol\ patch
would produce different measured lensing amplitudes. 
The cosmology dependence enters through the \none\ bias.
In this work, since we choose a fiducial cosmology that is consistent with the data, 
we expect the difference in the \none\ bias to be small. 
To test this, we sample the lensing amplitude given the fiducial cosmology
with and without corrections to \none. 
We find the difference in the lensing amplitude for the \mv\ estimator to be 0.007 (0.1\,$\sigma$).
We therefore neglect this correction in this work.

\begin{figure*}
\begin{center}
\includegraphics[width=0.98\textwidth]{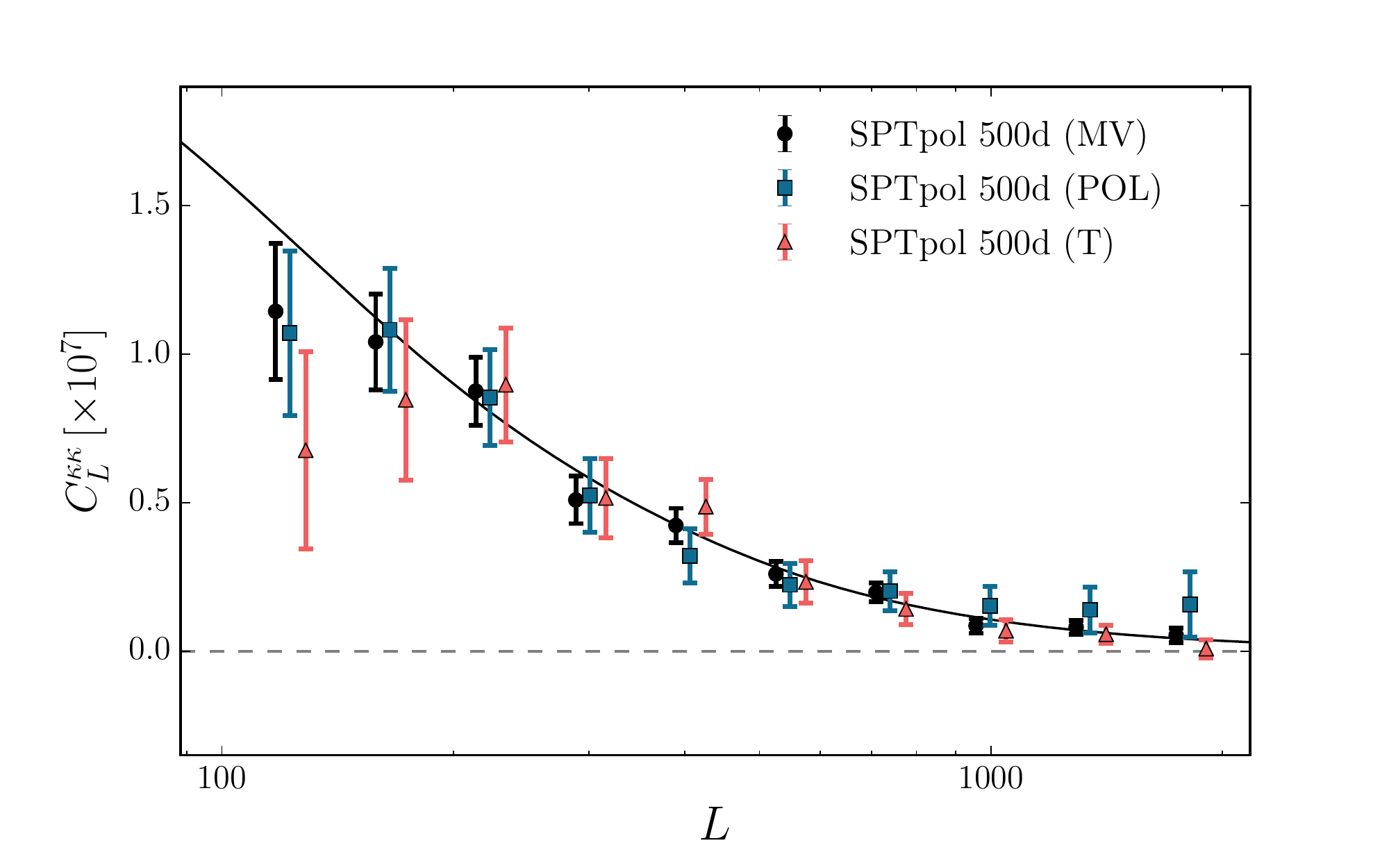}
\caption{Lensing convergence bandpowers estimated from \sptpol 500\,deg$^2$ field data.
We show bandpowers from the \mv, \pol, and \tt\ estimators. The \pol\ and \tt\ bandpowers are shifted in $L$ for clarity.
The \tt\ and \pol\ bandpowers are consistent with each other given the errorbars of the bandpowers.
The reconstruction noise of the \pol\ estimator is lower than that of \tt\ for $L \lesssim 600$, 
and vice versa on smaller angular scales.
This provides a sense of the angular scales at which each estimator gives better S/N. 
The black solid line shows the lensing convergence spectrum from the fiducial cosmology. }
\label{fig:main_bp}
\end{center}
\end{figure*}

\section{Results}
\label{sec:results}
\setcounter{footnote}{0}
In the following, we discuss our main results: the lensing convergence map, 
the lensing spectrum, and the lensing amplitude measurement. We compare our \mv\ map
to S15's. We discuss the relative weights of the \tt\ and the \pp\ results, 
compare the sizes of the 
systematic uncertainties to the statistical uncertainties, and 
put our measurement in the context of other lensing measurements.

In~\reffig{kappa_mv}, we show the \mv, \tt, and \pp\ lensing convergence.
The maps are smoothed by a 1-degree FWHM Gaussian to highlight the higher signal-to-noise
modes at the larger scales. 
At the angular scales shown, the \pp\ reconstruction has higher signal-to-noise (lower \nzero)
than the \tt\ reconstruction.
Therefore, while the \tt\ and the \pp\ $\kappa$ map fluctuations both trace those in 
the \mv\ map in broad strokes, we see that the \pp\ modes trace the \mv\ modes
more faithfully. 
The lensing modes in the \mv\ map are reconstructed with S/N better than unity for
$L \lesssim 250$. 
It is the largest lensing map reconstructed with this S/N level from the CMB to date.

In~\reffig{kappa_s15}, we compare the \mv\ $\kappa$ map from S15 
and the one from this work over the same region of the sky. 
We observe that the two convergence maps show nearly identical degree-scale structure. 
The S15 data were taken before the \sptpol\ 500\,deg$^2$ survey data (see~\refssec{survey}),
and thus constitute an independent dataset from that used in this work.
In addition, because the two analyses have similar 
per-mode reconstruction noise,
the visual agreement of the modes between the two 
is a useful consistency check. 

\begin{table}
\caption{\mv\ lensing bandpowers}
\centering
\begin{tabular}{c c c | c }
\hline\hline
$[\,L_{\rm min}$ & $L_{\rm max}\,]$ & $L_b$ & $10^7 \hat{C}_{L_{b}}^{\kappa\kappa}$ \\ [0.5ex]
\hline
$[\,100$&$133\,]$&$117$&$ 1.144\pm0.230$\\                                                                     
$[\,134$&$181\,]$&$158$&$ 1.041\pm0.161$\\                                                                     
$[\,182$&$244\,]$&$213$&$ 0.876\pm0.115$\\                                                                     
$[\,245$&$330\,]$&$288$&$ 0.509\pm0.080$\\                                                                     
$[\,331$&$446\,]$&$389$&$ 0.423\pm0.058$\\                                                                     
$[\,447$&$602\,]$&$525$&$ 0.260\pm0.042$\\                                                                     
$[\,603$&$813\,]$&$708$&$ 0.198\pm0.032$\\                                                                     
$[\,814$&$1097\,]$&$956$&$ 0.086\pm0.025$\\                                                                    
$[\,1098$&$1481\,]$&$1290$&$ 0.081\pm0.023$\\                                                                  
$[\,1482$&$1998\,]$&$1741$&$ 0.053\pm0.025$\\    
\hline
\end{tabular}
\tablecomments{The bandpowers for the \mv\ spectrum 
as defined in~\refeq{bandpowers} and shown in~\reffig{main_bp}.
Bins are evenly spaced in $\log(L)$ and bandpowers are reported at the center of each bin.}
\label{tab:bandpowers}
\end{table}

We present the lensing power spectrum measurement in logarithmically spaced bins
in the range $100 < L < 2000$. 
We list the \mv\ lensing bandpowers and their uncertainties in~\reftab{bandpowers}.
The lower bound of the $L$ range is chosen by the region of validity of the MC bias correction.
The upper bound is set by computing the uncertainties on the lensing amplitudes
as we include higher multipoles and seeing no gain in signal-to-noise of the amplitude
going beyond $L$ of 2000. 
In~\reffig{main_bp}, we show the bandpowers from the \mv\ spectrum, the \pp\ spectrum, and 
the \tt\ spectrum.
We see that the error bars of the \pp\ spectrum are smaller than those of the \tt\ spectrum at
$L \lesssim 600$, and vice versa on smaller angular scales.
This is consistent with the \nzero\ of the \tt\ reconstruction at small angular scales being lower than that of
the \pp\ reconstruction -- the small angular scale modes are better reconstructed
by the \tt\ estimator. 

\begin{figure*}
\begin{center}
\includegraphics[width=0.98\textwidth]{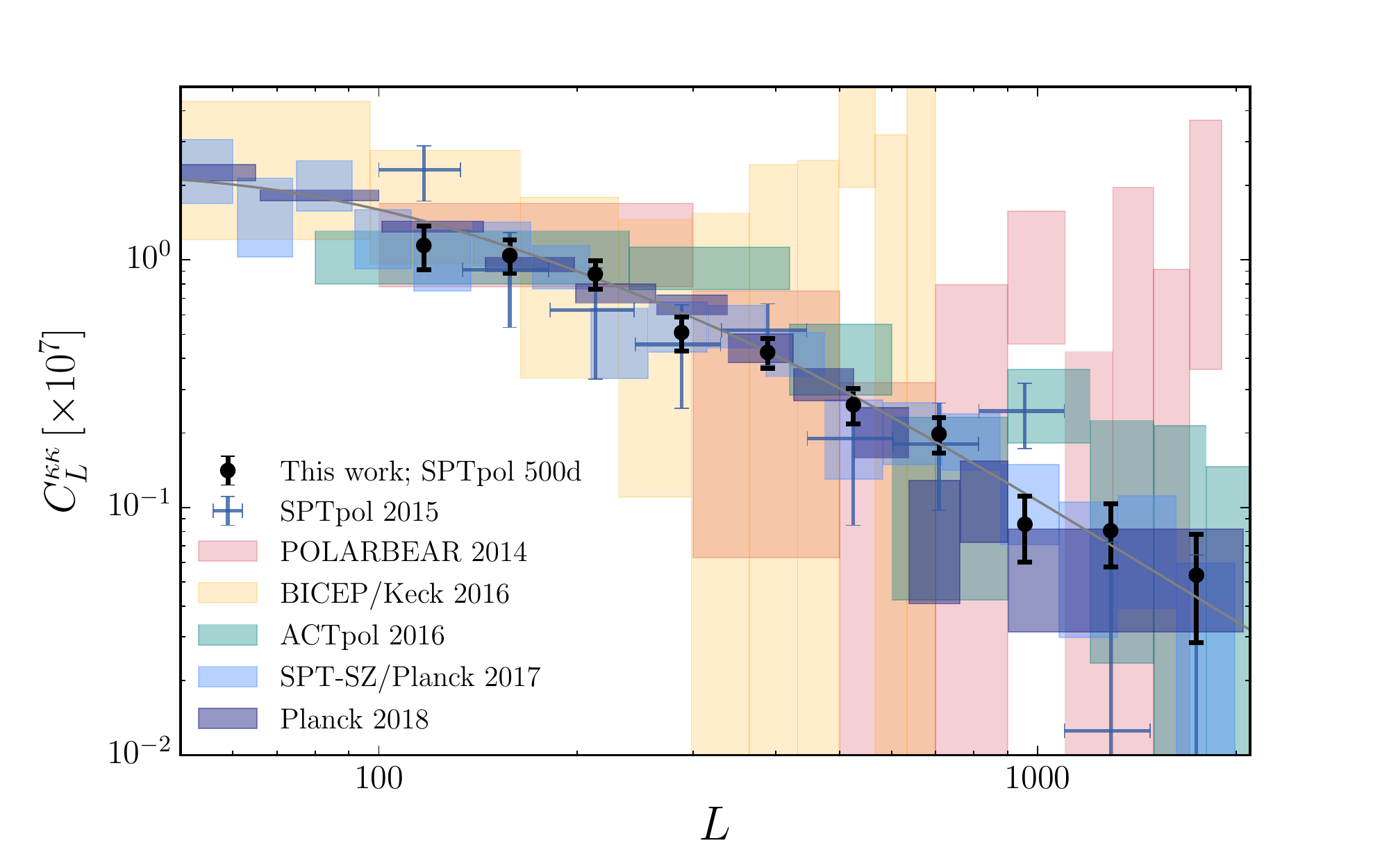}
\caption{Lensing power spectrum measurements from this work and S15 (\mv)~\citep{story14}, 
POLARBEAR (\pp)~\citep{polarbear2014a}, BICEP2/Keck (\pp)~\citep{bicep2keck16},
ACTPol (\mv)~\citep{sherwin16}, SPT-SZ + \planck (\tt)~\citep{omori17}, and \planck (\mv)~\citep{planck18-8}.
The gray solid line is the lensing spectrum from 
the best-fit \LCDM model to the \planck~\texttt{plikHM\_TT\_lowTEB\_lensing} dataset.
This work contains data from $\sim$5 times more area than S15 and the sizes of the 
errorbars of the signal-dominated multipoles reflect the decrease in sample variance. 
In addition, it is the tightest lensing amplitude measurement of all the lensing measurements made using
only ground-based telescope data to date.}
\label{fig:global}
\end{center}
\end{figure*}

We measure the overall lensing amplitude for each estimator and find
\begin{multline*}
\begin{aligned}
A_{\rm\mv} &=  \mvAmp \pm \mvAmpStat {\rm\,(Stat.)}\pm\mvAmpSys{\rm\,(Sys.)}\,, \\
A_{\rm\pp} &= \ppAmp \pm \ppAmpStat {\rm\,(Stat.)}\pm\ppAmpSys{\rm\,(Sys.)} \,, \\
A_{\rm\tt} &= \ttAmp \pm \ttAmpStat {\rm\,(Stat.)}\pm\ttAmpSys{\rm\,(Sys.)}  \,.
\end{aligned}
\end{multline*}
We derive the statistical uncertainties from the standard deviations of the lensing amplitudes from simulations.
We detail the sources of systematic uncertainties in~\refssec{sys} and break them down in~\reftab{sys_uncer}. 
Considering statistical uncertainties alone, 
we measure the lensing amplitude with \mvPercent\ uncertainty using
the \mv\ estimator and with \ppPercent\ uncertainty with the \pp\ estimator.
For the \tt\ estimator, we measure the lensing amplitude with \ttPercent\ uncertainty.
Having chosen the same cuts in multipole space for both the
input temperature and polarization maps, 
this shows that the signal-to-noise per mode in the input polarization maps are now high
enough that the \pp\ estimators give more stringent measurements
of the lensing amplitude than the \tt\ estimator. 
In future analyses, the \tt\ lensing spectrum is sample-variance limited and cannot
be improved by lowering the temperature map noise levels. 
Instead, it can be improved by including information from 
higher multipoles and/or more sky area.
However, lowering the noise levels of the polarization maps can still improve
the lensing measurement from polarization estimators. 
Specifically, unlike the temperature estimator, the \nzero\ of the $EB$ estimator is not limited by 
unlensed power in the map, because there is little unlensed $B$ mode power to contribute
to \nzero\ in the multipole range important for lensing reconstruction.
In addition to surpassing the measurement uncertainty of the \tt\ lensing amplitude,
considering statistical uncertainties alone,
our \pp\ lensing amplitude is the most precise amplitude measurement (\ppLensSig)
using polarization data alone to date. 

The systematic uncertainties for the \mv\ and the \pp\ estimators 
are $\sim$40\% of their respective statistical uncertainties, 
whereas the systematic uncertainty is subdominant
for the \tt\ estimator compared with its statistical uncertainty. 
For both the \mv\ and the \pp\ estimators, 
the systematic uncertainty budget is dominated by the \pcal\ uncertainty
(\refssec{sys}).
Including the systematic uncertainties in the \mv\ amplitude measurement, we measure
$A_{\rm \mv}$ with \mvTotPercent\ uncertainty. 

We detect lensing at very high significance. 
From reconstructing $\phi$ using 400 unlensed simulations, 
the standard deviation of  $A^{\rm unl}_{\rm \mv}$ is 0.024.
The observed amplitude of $A_{\rm \mv}=\mvAmp$ would thus correspond to a $\mvNullSig$ fluctuation.

Compared with other ground-based measurements, our result has the tightest constraint 
on the lensing amplitude.
In~\reffig{global}, we show our lensing power spectrum measurement
against previous measurements.  
Our measurement is consistent with the measurement by \cite{omori17}. 
In that work, they reconstruct lensing using a combined temperature map 
from SPT-SZ and \planck\ over the common 2500\,deg$^2$ of sky. 
They measure the lensing amplitude to be $0.95 \pm 0.06$ relative to 
the best-fit \LCDM\ model to the \planck 2015 \texttt{plikHM\_TT\_lowTEB\_lensing} dataset
(same as the fiducial cosmology used in this work). 
The most recent lensing analysis of all-sky \planck\ data found the best-fit lensing amplitude
to be $1.011 \pm 0.028$ against the \planck\ 2018 \texttt{TTTEEE\_lowE\_lensing}
cosmology~\citep{planck18-8}.
To compare our measurement with this model, we refit our minimum-variance bandpowers and
get $A_{\rm\mv} = 0.946 \pm 0.058 {\rm\,(Stat.)} \pm \mvAmpSys {\rm\,(Sys.)}$, consistent
with \planck's lensing measurement.

\section{Null Tests, Consistency Checks, and Systematic Uncertainties}
\label{sec:tests}
In this section, we summarize the null and consistency tests we have performed on the data and account 
for the systematic uncertainties in our lensing amplitude measurements. 
We report test results from the \mv, \pp, and \tt\ estimators. 

\subsection{Null Tests and Consistency Checks}
\label{ssec:null}
\begin{figure}
\begin{center}
\includegraphics[width=0.48\textwidth]{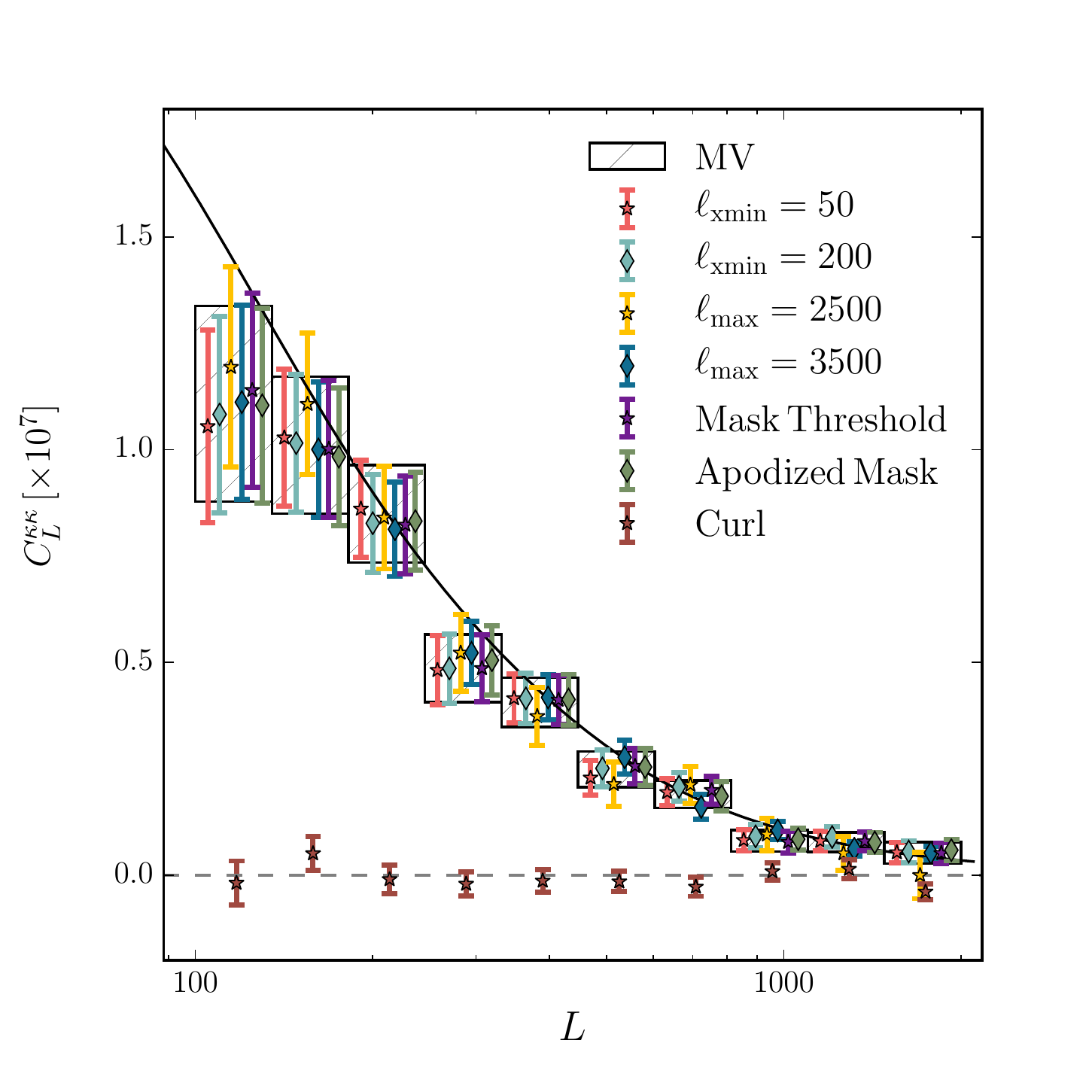}
\caption{Results of the power spectrum consistency tests and curl null test. 
The figure shows $C_L^{\kappa\kappa}$ bandpowers from different analysis choices
and the curl test bandpowers. 
The \mv\ bandpowers and errorbars from the baseline analysis are plotted as boxes. 
The bandpowers calculated with different analysis choices are consistent
with the baseline \mv\ bandpowers.
The bandpowers from the curl null test are consistent with zero.
}
\label{fig:consistency_bandpower}
\end{center}
\end{figure}

We quantify the results of our tests using summary statistics comparing data bandpower differences 
with simulation bandpower-difference distributions,
where the differences are taken between bandpowers obtained from the baseline analysis pipeline
and from a pipeline with one analysis change.
We plot these bandpowers
in~\reffig{consistency_bandpower} to provide an absolute sense 
of how much the bandpowers and their error bars change
given the various analysis variations.

\begin{figure*}
\begin{center}
\includegraphics[width=0.98\textwidth]{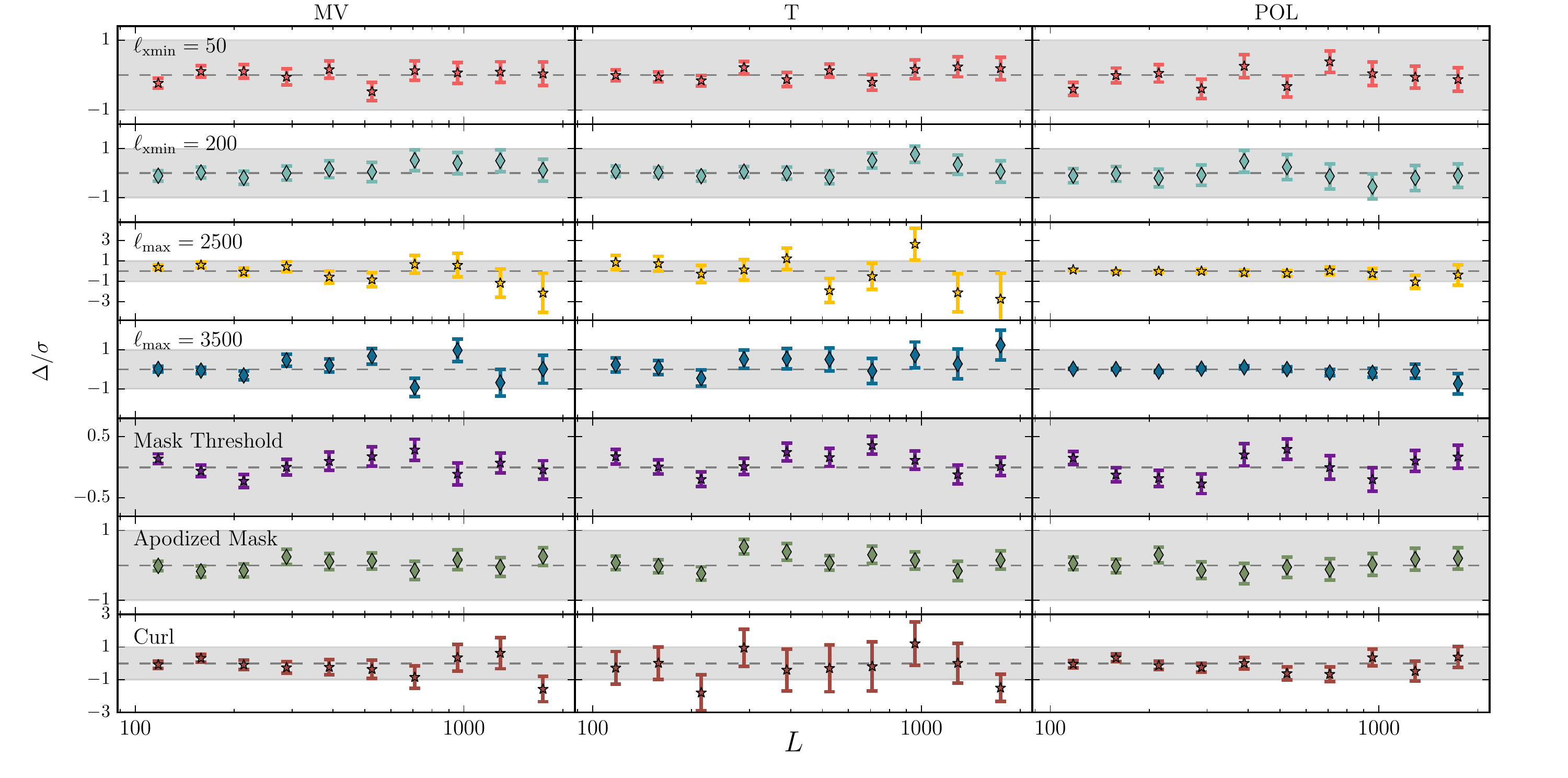}
\caption{The difference bandpowers ($\Delta C_L^{\kappa\kappa}$) and their 
uncertainties between the baseline analysis and analyses with one change
as indicated in the labels,
scaled by the uncertainties 
of the respective \mv, \tt, \pp\ bandpowers.
The 1\,$\sigma$ bands of the \mv, \tt, and \pp\ estimators are shaded in gray. 
This is a visual summary of the $\chi^2$ values listed in~\reftab{null_pte}.
We find the analysis-variation bandpowers and best-fit lensing amplitudes to be consistent
with the baseline setup for all three estimators. 
}
\label{fig:delta_sigma}
\end{center}
\end{figure*}

Quantitatively, we calculate the $\chi^2$ of the data-difference spectrum
$(\Delta C^{\phi \phi}_{b,\,{\rm data}})$ against 
the mean of the simulation-difference spectrum $(\langle \Delta C^{\phi \phi}_{b,\,{\rm sim}}\rangle)$
using the variance of the simulation-difference spectra ($\sigma_{b,\,{\rm sys}}^2$),
as the difference bandpowers are largely uncorrelated: 
\beq
\label{eq:chi2}
\chi^2_{\rm sys} = \sum_b{ \frac{(\Delta C^{\phi \phi}_{b,\,{\rm data}} - \langle \Delta C^{\phi \phi}_{b,\,{\rm sim}}\rangle)^2}
                                { \sigma_{b,\,{\rm sys}}^2} } \,.
\eeq
The probability-to-exceed (PTE) is calculated from a $\chi^2$ distribution with
10 degrees of freedom, as we have 10 bandpowers. 
For the curl test, $\sigma_{b,\,{\rm sys}}$ comes from the distribution of the simulation
curl spectra. 
\reffig{delta_sigma} provides a visual summary of these tests. 
It shows the data-difference bandpowers for each test for the
three estimators. 
The error bars are generated from the distributions of simulation-difference bandpowers.
The data points and the error bars from each estimator in each $L$ bin 
are scaled by the 1\,$\sigma$ lensing spectrum
uncertainties from that estimator in that bin.  
We list the $\chi^2$ and PTE values for each test and each estimator in \reftab{null_pte}.

We compare the differences in overall lensing amplitude similarly.
We calculate the difference in the lensing amplitudes
between the baseline and the alternate analysis choice setup:  $\Delta A_{\rm data}$.
We form the $\chi^2$ by comparing the data-difference amplitude against the distribution
of the difference amplitudes in simulations. \\

\begin{table*}
\caption{Consistency and Null tests $\chi^2$ and PTE}
\centering
\begin{adjustbox}{max width=1\textwidth}
\begin{tabular}{l |c c|c c | c c|c c | c c|c c }
\hline\hline
Test Name & $\chi^2_{\rm\mv}$ & (PTE)  & $\Delta$ $A_{\rm \mv}$ & (PTE)
                  & $\chi^2_{\rm\tt}$ & (PTE)  & $\Delta$ $A_{\rm \tt}$ & (PTE)
                  & $\chi^2_{\rm\pol}$ & (PTE)  & $\Delta$ $A_{\rm \pol}$ & (PTE)\\ 
		 &&& $\pm {\rm var}(\Delta A_{\rm \mv})$&      
		&&& $\pm {\rm var}(\Delta A_{\rm \tt})$&   
		&&& $\pm {\rm var}(\Delta A_{\rm \pol})$&   \\ [0.5ex]
\hline
$\ell_{\rm xmin}$=50 & 7.2 & (0.71) & -0.002 $\pm$ 0.013 & (0.76)
      & 5.8 & (0.83) & 0.005 $\pm$ 0.019 & (0.86)
      & 9.7 & (0.47) & -0.015 $\pm$ 0.025 & (0.55) \\
$\ell_{\rm xmin}$=200 & 4.3 & (0.93) & 0.017 $\pm$ 0.021 & (0.44)
      & 10.4 & (0.40) & 0.035 $\pm$ 0.026 & (0.17)
      & 3.3 & (0.97) & -0.005 $\pm$ 0.037 & (0.90) \\
$\ell_{\rm max}$=2500 & 14.2 & (0.17) & -0.000 $\pm$ 0.035 & (0.98)
      & 12.0 & (0.28) & 0.040 $\pm$ 0.093 & (0.65)
      & 5.9 & (0.83) & -0.021 $\pm$ 0.018 & (0.25) \\
$\ell_{\rm max}$=3500 & 17.2 & (0.07) & 0.014 $\pm$ 0.023 & (0.66)
      & 8.2 & (0.61) & 0.085 $\pm$ 0.055 & (0.17)
      & 11.0 & (0.36) & -0.006 $\pm$ 0.008 & (0.48) \\
Mask thres.& 13.2 & (0.21) & 0.006 $\pm$ 0.008 & (0.43) 
      & 15.6 & (0.11) & 0.021 $\pm$ 0.014 & (0.10) 
      & 14.6 & (0.15) & -0.002 $\pm$ 0.013 & (0.88)\\
Apod. mask & 4.8 & (0.91) & 0.004 $\pm$ 0.012 & (0.90)
      & 13.3 & (0.20) & 0.035 $\pm$ 0.022 & (0.13)
      & 3.7 & (0.96) & -0.002 $\pm$ 0.022 & (0.85) \\
Curl & 9.3 & (0.50) & -0.007 $\pm$ 0.005 & (0.20)
& 7.5 & (0.67) & -0.013 $\pm$ 0.013 & (0.29)
& 9.4 & (0.49) & -0.010 $\pm$ 0.009 & (0.28)\\
\hline
\end{tabular}
\end{adjustbox}
\tablecomments{Results of systematics tests. 
For each test, we show the $\chi^2$ and PTE of the difference bandpowers
and the difference amplitudes and its associated PTEs 
for each of the \mv, \tt, and \pp\ estimators. 
}
\label{tab:null_pte}
\vspace{0.1cm}
\end{table*}
 
{\it Varying $\ell_{\rm xmin}$ and $\ell_{\rm max}$:} 
We vary the multipole range of the
input CMB maps used for lensing reconstruction to check 
the following: 
(1) consistency of bandpowers as we include more or fewer CMB modes and
(2) impact of foregrounds as we increase the maximum multipole.
As we increase the maximum multipole used in the input temperature map, 
the contamination of the CMB by foregrounds like tSZ and CIB increases.
As discussed in~\refssec{lensspec}, both of these inputs can bias the lensing spectrum.
On the low $\ell$ side, we remove $|\bl_{x}| < \ell_{\rm xmin}$ modes because the combination of our
observing strategy and time stream filtering removes
modes approximately along the $\ell_{x} = 0$ axis.
We run two cases for the $\ell_{\rm xmin}$ cut: $\ell_{\rm xmin} = 50$ and 
$\ell_{\rm xmin} = 200$. 
From ~\reftab{null_pte}, we see that the differences of the $\ell_{\rm xmin}$ cases 
are consistent with the expected simulation distribution. 
We can in principle set the $\ell_{\rm xmin}$ in the baseline analysis to be 50. 
However, since the applied time stream filtering removes $\ell \lesssim 100$ modes, 
and the number of modes available between $\ell_{\rm xmin}$ of 50 and 100 is small
compared with the number of modes at the high-$\ell$ end,
we do not expect there to be significant improvement in signal-to-noise and therefore
choose to set $\ell_{\rm xmin} =100$ for the baseline analysis.
For the $\ell_{\rm max}$ tests, we vary the $\ell_{\rm max}$ cut between $\bl = 2500$
and $\bl = 3500$ in steps of 500. 
We observe from~\reftab{null_pte} that moving between $\ell_{\rm max}$ of 3000 and 
2500 and between $\ell_{\rm max}$ of 3000 and 3500, the data bandpower differences
are consistent with bandpower differences from simulations for all three sets of estimators.
We set the baseline $\ell_{\rm max}$ to be 3000 to keep the systematic uncertainty due 
to the subtraction of foreground biases subdominant to the statistical uncertainty of the measurements. 
For future analyses, estimators designed to reduce foreground 
biases~\citep{osborne14, schaan18, madhavacheril18}
can be employed, potentially sacrificing some statistical power for reduced systematic uncertainty. \\

\begin{figure*}
\begin{center}
\includegraphics[width=0.98\textwidth]{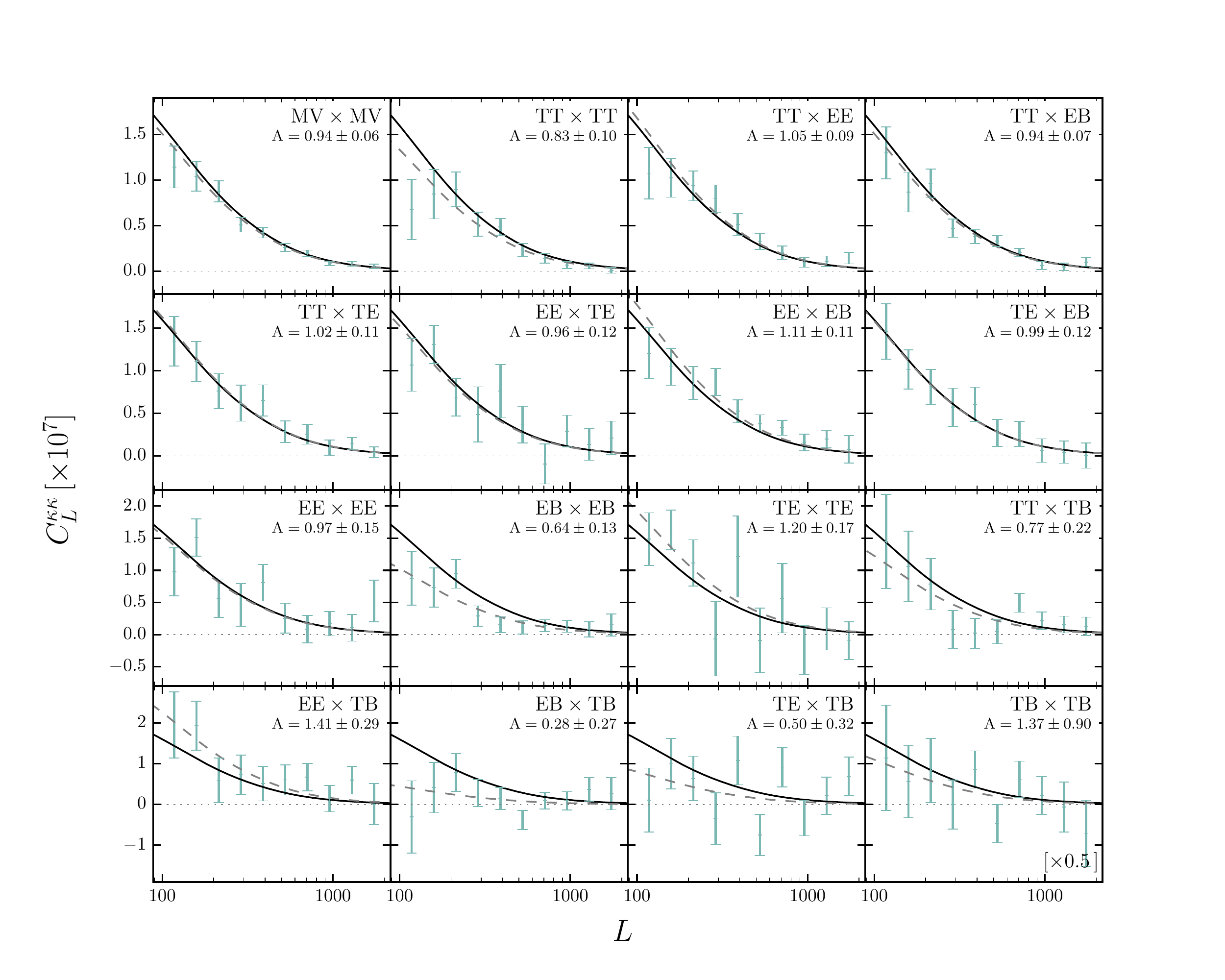}
\caption{Comparison of cross-spectra of individual estimators and the  minimum variance combination \mv$\times$\mv.
The lensing spectrum from the fiducial cosmology, the \planck~\texttt{plikHM\_TT\_lowTEB\_lensing} 
\LCDM\ best fit, is plotted in solid black. The best fit from each estimator is plotted in dashed gray. 
The amplitudes are computed relative to the fiducial lensing spectrum. 
The data points and errorbars are scaled by 0.5 for the TB$\times$TB spectrum to allow all the points to be shown.
We find a ${\rm PTE}=0.44$ when comparing the 150 bandpowers 
against the best fit of the \mv\ spectrum using a covariance matrix constructed from simulations.}
\label{fig:16_panel}
\end{center}
\end{figure*}

{\it Source masking threshold:}
We check the impact of extragalactic foregrounds on the lensing measurements
by varying the masking thresholds of point sources and galaxy clusters.
We raise the thresholds of the point source and the cluster masks
so that about 20\% more sources are kept in the input maps.
The modified cuts correspond to a flux threshold of 7.5 mJy and 
a cluster detection significance threshold of 4.7, 
whereas the baseline thresholds (see~\refssec{data}) are 6 mJy and 4.5 
(i.e., the threshold for inclusion in the \citealt{bleem15} catalog).
From the bandpower PTEs and amplitude PTEs listed in~\reftab{null_pte},
we see that the differences in bandpowers and amplitudes are consistent
with expected variations from simulations. 
Therefore, we conclude that the masking thresholds we apply in the baseline
analysis reduce foreground biases to the lensing spectrum sufficiently. 
\\

{\it Apodization:} 
We test for the effects of using a top hat function 
for the boundary and source mask by
redoing the entire analysis using a cosine taper instead.
The radius of the cosine taper is 10$^{\prime}$ for the boundary mask
and 5$^{\prime}$ for the sources. 
The data difference is typical given the simulation-difference distribution for
all three estimators. \\

{\it Curl test:} The deflection field $\bold{d}(\hat{\bold{n}})$ that remaps the primordial CMB anisotropies
can be generically decomposed into a gradient and a curl component: 
\beq
\label{eqn:curl}
\bold{d}(\hat{\bold{n}}) = \nabla \phi(\hat{\bold{n}}) + \star\nabla \Omega(\hat{\bold{n}}) \,,
\eeq
where $\star$ is a 90$^{\circ}$ rotation operator and $\Omega(\hat{n})$ is a 
psuedo-scalar field that sources the curl component~\citep[see e.g.][]{hirata03b, namikawa12}.
In the Born approximation, the lensing potential only sources the gradient component
of the deflection field.
At current noise levels and barring unknown physics, the curl component is expected
to be consistent with zero. 
Therefore, estimating the curl of the deflection field serves as a test of
the existence of any components in the data generated from non-Gaussian
secondary effects or foregrounds~\citep{cooray05}.
Analogous to estimating the lensing $\phi$ field, for the curl component, 
we estimate the psuedo-scalar field $\Omega$. 
The curl estimator is orthogonal to the lensing (gradient) estimator and
has the same form as the lensing estimator~\refeq{phi_bar}.
The weights of the curl estimator are designed to have response to the curl 
instead of the gradient. 
We implement the weights as presented in~\citet{namikawa12}. 
The curl spectrum is derived analogous to the lensing spectrum with two differences:
(1) the theory input is set to a flat spectrum $C_L^{\Omega\Omega} = 10^{-7}$, where
it is used for uniformly weighting the modes when binning and as a reference 
spectrum for the amplitude calculation; 
(2) no response correction from simulations is applied to the psuedo-scalar field as the 
expected signal is zero. 
The bottom panel of~\reffig{delta_sigma} shows the power spectra for the \mv, \pp,
and \tt\ curl estimators and their uncertainties scaled by the respective uncertainties
of their lensing spectra.  
The bandpower PTEs and the amplitude PTEs for the three estimators are listed in~\reftab{null_pte}.
They are all consistent with zero.
The bandpower PTE and the amplitude PTE are 0.50 and 0.20 for the \mv\ curl estimator. 
We thus see no evidence of contamination to our lensing estimate from non-Gaussian secondaries.\\

{\it Consistency across estimators:}
In~\reffig{16_panel}, we show the reconstructed lensing spectrum from all 15 
pairs of the 5 estimators $TT, TE, EE, EB, TB$, and the MV spectrum.
We test for consistency of the lensing spectra from the 
15 pairs of estimators with the best fit from the \mv\ estimator
by calculating $\chi^2$ and PTE.
We calculate the $\chi^2$ of the 150 bandpowers against the 
binned theory spectrum scaled by the best-fit \mv\ amplitude:
\beq
\chi^2 = (d - m) \, {\rm Cov}^{-1} (d - m)^{\dagger}\,,
\eeq
where $d$ is the 150 data bandpowers, $m$ is 15 copies of the scaled binned theory spectrum,
and ${\rm Cov}$ denotes the covariance matrix. 
We construct the covariance matrix of the 150 bandpowers of 15 estimators
from 400 simulations. 
We set the off-diagonal terms of each subblock off the main diagonal to zero as we expect there to be
and have verified that there is little
to no correlation across different $L$ bins between estimators.
For the main diagonal, the first and second bins are correlated at the $\sim$10\%
level for 12 of the estimators, and we keep them while setting the rest of the covariance
elements to zero. 
The PTE compared to the set of 400 sets of simulated bandpowers is 0.44. 
Comparing instead the best-fit amplitudes of the 15 pairs of estimators with the \mv\ 
best-fit amplitude,
we note that the EB$\times$TB and EB$\times$EB pairs
are low by 2.4\,$\sigma$ 
and 2.1\,$\sigma$ respectively, while 
the rest of the pairs are within 2\,$\sigma$ of the \mv's amplitude. 
With the number of tests we have performed, it is not unusual to see 2\,$\sigma$ outliers.

\subsection{Systematic Uncertainty}
\label{ssec:sys}
In this section, we summarize the sources of systematic uncertainty and our
accounting of them in the lensing amplitude measurement.
We quantify the uncertainties in the lensing amplitude measurement due to uncertainties
in the beam measurement, temperature and polarization calibrations, 
T$\rightarrow$P leakage correction, global polarization angle rotation,
and the applied foreground template. 
We address the potential impact of non-Gaussian polarized foregrounds. 
The sources of systematic uncertainty and their respective impact on the lensing amplitude
measurements are listed in~\reftab{sys_uncer}. \\

{\it Beam uncertainty:} We take the beam measurement uncertainty $\Delta B_{\ell}$
derived from the beam covariance matrix in H18 and convolve (1+ $\Delta B_{\ell}$) 
with the input $\bar{X}$ maps while keeping all the simulation maps the same.
We analyze the data maps as if their beam were the mean measured beam $B_{\ell}$
as opposed to the modified $B_{\ell} (1+\Delta B_{\ell})$. 
The difference in the lensing amplitude between this analysis and the baseline analysis
quantifies the effect of an underestimation of the beam profile by 1\,$\sigma$ of the
beam measurement uncertainty across the entire angular multipole range.
We find $\Delta A_{\rm\mv} = \mvSysBeam$, $\Delta A_{\rm\pp} = \ppSysBeam$, and 
$\Delta A_{\rm\tt}=\ttSysBeam$. 
The systematic uncertainties induced by the beam uncertainty are small ($\sim$0.1\,$\sigma$) 
compared with the respective statistical uncertainties 
of the measured lensing amplitudes in \tt, \pp, and \mv. \\

{\it Temperature and polarization calibrations:} 
We apply the temperature and polarization calibration factors derived in H18
to our data $T/Q/U$ maps as described in~\refssec{data}. 
From the posterior distributions of H18, we obtain
the uncertainties of the \tcal\ factor $\delta$\tcal\ and 
the \pcal\ factor $\delta$\pcal\, which are 0.3\% and 0.6\% respectively. 
The \tcal\ and \pcal\ are applied to the raw temperature and polarization maps
through $T = T^{\rm raw} \times$~\tcal\ and 
$(Q/U) = (Q/U)^{\rm raw} \times$ \tcal $\times$ \pcal.  
Keeping the simulated maps fixed to the baseline analysis, we scale the
data maps by $(1+\delta$\tcal) for the temperature map and 
(1+$\delta$\tcal)(1+$\delta$\pcal) for the polarization maps and we recalculate
the data lensing amplitudes\footnote{In S15, we accounted for the systematic uncertainties 
from calibration uncertainty as 4$\times \delta \tcal$ and 4$\times \delta \pcal$.
This underestimates the systematic uncertainty because the object that gets scaled by
$\delta \tcal$ and $\delta \pcal$ is $\cLppHat$ instead of $\hatcLpp$.
This means that the calibration offset applies to the noise biases terms as well as the
signal term and can be a factor of a couple larger than $\hatcLpp$ itself.
}.
The two pieces in the amplitude calculation that change because of
 the modified data maps are  $C^{\hat{\phi} \hat{\phi}}_{L}$ and \rdn.
To illustrate how they contribute to the overall change in the
lensing amplitude, we consider here the temperature-only estimator. 
$C^{\hat{\phi} \hat{\phi}, \tt}_{L}$, which contains four copies of the data map,  
would shift by $(1+\delta \tcal)^4$. 
For \rdn, the four terms where two of the four input maps are replaced by 
data maps get a multiplicative correction of $(1+\delta \tcal)^2$.  
The overall shift in $\hat{C}^{\phi\phi, \tt}_{L}$ is therefore
\beq
\label{eq:sys_calib} 
\Delta \hatcLpp  = \Delta \cLppHat - \Delta \rdn \,,
\eeq
where 
\beq
\Delta \cLppHat  =  [(1+\delta \tcal)^4 -1]  \cLppHat  
                   \simeq   4 \delta \tcal \, \cLppHat 
\eeq
and

\begin{multline}
\begin{aligned}
 \Delta \rdn  \simeq
        2 \delta & {\rm T_{cal}} \,  
 \big \langle  \, \cLppHat[\bar{T}_{\rm d},  \bar{T}_{\rm MC},  \bar{T}_{\rm d},   \bar{T}_{\rm MC}] \\
    &\quad +\cLppHat[\bar{T}_{\rm MC}, \bar{T}_{\rm d},   \bar{T}_{\rm d},   \bar{T}_{\rm MC}]  \\
  &\quad +\cLppHat[\bar{T}_{\rm d},  \bar{T}_{\rm MC},  \bar{T}_{\rm MC},  \bar{T}_{\rm d} ] \\
  & \quad   +\cLppHat[\bar{T}_{\rm MC}, \bar{T}_{\rm d},   \bar{T}_{\rm MC},  \bar{T}_{\rm d} ]  \,
  \big \rangle_{\rm MC} \, ,
 \\
\end{aligned}
\end{multline}
with ${\rm MC}$ denoting simulation realizations and $d$ denoting data. 
The polarization-only case has a similar form as the temperature-only case
with $(1+\delta\tcal) \rightarrow (1+\delta\tcal)(1+\delta\pcal)$.
The difference in the resultant measured amplitudes due to an offset in the
calibration factors depends on the relative amplitudes of 
$\cLppHat$ and \rdn\ at each multipole $L$. 
Therefore, we quantify the difference by running the baseline analysis 
with the temperature and polarization calibration of the data maps shifted by 1\,$\sigma$.
We find that $\Delta A_{\rm \mv} =\mvSysCal$, $\Delta A_{\rm \pp} = \ppSysCal$, and 
$\Delta A_{\rm \tt}=\ttSysCal$.
The shifts in the lensing amplitudes are dominated by the polarization calibration uncertainty. 
Furthermore, the calibration systematic uncertainty is  
almost half of the statistical uncertainty of the polarization-only amplitude.
Since the signal-to-noise of future lensing measurements will be driven by the polarization estimators, 
we will need more precise polarization calibration in order for the measurements to remain dominated by
statistical uncertainties. 
This can be achieved by cross-calibrating with deeper CMB maps or over larger areas of sky,
assuming an external CMB map exists with more accurate polarization calibration than \planck.
Relatedly, lensing measures mode coupling 
which in principle can be extracted irrespective of the input maps calibration.
An example of circumventing the systematic uncertainty contribution caused by 
calibration uncertainties in the input maps is to use the measured power spectra
in one of the weight functions in the analytic response in~\refeq{phi_resp}.
In this approach, the response moves together with the calibration of the input maps
and therefore eliminates this systematic uncertainty.\\

{\it Temperature-to-polarization leakage:}
We estimate the bias to the lensing amplitude measurement caused by misestimating 
the T$\rightarrow$P leakage factors.
Similar to quantifying the systematic offsets by the beam uncertainty and the \tcal/\pcal\ 
uncertainties, here 
we modify the data polarization maps by over-subtracting a $\epsilon^{(Q/U)}$-scaled
copy of the temperature map by 1\,$\sigma$ in $\epsilon^Q$ and $\epsilon^U$, 
while keeping the rest of the analysis the same. 
We find the change in $A_{\rm \mv}$ and $A_{\rm \pp}$ to be negligible, at
0.001\,$\sigma$ and 0.002\,$\sigma$ respectively.  \\

{\it Global polarization angle rotation:}
There is a 6\% uncertainty in the global rotation angle measured through the minimization
of the $TB$ and $EB$ spectra discussed in~\refssec{data}. 
How much would an offset in the polarization angle rotation bias the lensing amplitude
measurement? 
We run the baseline analysis with the data polarization maps rotated an extra 6\% on top
of the measured angle from the minimization procedure. 
We find that $A_{\rm \mv}$ and $A_{\rm \pp}$ change by less than 0.01\,$\sigma$ of their
respective statistical uncertainties. \\

{\it Extragalactic foregrounds:}
As discussed in~\refssec{lensspec}, we subtract templates of expected foreground biases 
from the \tt\ and \mv\ lensing power spectra $ \hatcLpp$ 
given models of CIB and tSZ from \cite{vanengelen14a}. 
The templates are taken from the mean of the range of models.
We estimate the systematic uncertainty from the template subtraction step 
by measuring the lensing amplitude difference using the maximum and minimum of 
template values allowed by the model space. 
We find the lensing amplitudes $A_{\rm \mv}$ to shift by $\pm\mvSysFg$ and
$A_{\rm \tt}$ to shift by $\pm\ttSysFg$, both less than 0.1\,$\sigma$ of their statistical 
uncertainties. 
For the temperature estimator, this source of bias is $\sim$1\% of the lensing amplitude
and is of the same magnitude as the bias from the temperature calibration.  
For polarization, we expect the only significant source of extragalactic foregrounds 
to come from the unclustered point-source component, 
because SZ and the clustered CIB are negligibly polarized.
The SZ effects are expected to be polarized at the 
less than 1\% level for the cluster masking threshold of this work~\citep{birkinshaw99,carlstrom02}.
The clustered CIB component is a modulation of the mean power 
from all sources~\citep{scott99} and is thus effectively unpolarized.
For the unclustered point sources, we measure the mean squared polarization
fraction to be 
$\langle p^2 \rangle = (\, 0.95 \pm 0.11 \,) \times 10^{-3}$ ($\sim$3\% polarized) at 150\,GHz
for sources above 6\,mJy in the \sptpol\ field~\citep{gupta19}. 
Assuming the polarization fraction is constant with flux density, 
the extragalactic foregrounds are thus on average significantly less polarized than the CMB. 
Since the level of bias introduced from temperature foregrounds is at the
percent level, we conclude that
the bias from polarized sources is negligible for this work. \\

\begin{table}
\caption{Systematic Uncertainties}
\centering
\begin{tabular}{l | r | r | r}
\hline\hline
Type & $\Delta A_{\rm \mv}$ & $\Delta A_{\rm \pp}$ & $\Delta A_{\rm \tt}$  \\ [0.5ex]
\hline
$\Delta A_{\rm beam}$       & \mvSysBeam   & \ppSysBeam & \ttSysBeam  \\
$\Delta A_{\rm cal}$      & \mvSysCal  & \ppSysCal   & \ttSysCal  \\ 
$\Delta A_{\rm T\rightarrow P}$  &      $\ll0.001$     & $\ll 0.001$  &  N/A  \\
$\Delta A_{\rm pol. rot.}$  &   $<0.001$  & $< 0.001$  &  N/A  \\
$\Delta A_{\rm fg}$        &  \mvSysFg  & N/A  & \ttSysFg \\
\hline
$\Delta A_{\rm tot}$       & \mvAmpSys  & \ppAmpSys & \ttAmpSys 
\end{tabular}
\tablecomments{The contributions to the systematic uncertainty budget. 
The total systematic uncertainties are evaluated by taking the quadrature sum of
the individual contributions.}
\label{tab:sys_uncer}
\end{table}

{\it Galactic foregrounds:}
Galactic foregrounds are non-Gaussian and can contribute to both the lensing estimator,
imparting a bias to the lensing spectrum measurement, and the curl estimator, 
causing the curl null test to fail. 
We mitigate these potential effects by (1) observing a low-foreground patch, 
(2) suppressing power from angular scales that are foreground dominated.
We use the curl null test to check for potential contamination from galactic foregrounds. 
The \sptpol 500\,deg$^2$ field is one of the lowest galactic foreground regions~\citep[e.g.][]{planck18-11}.
Galactic foreground power is fainter than the CMB in both temperature and in 
$E$-mode polarization for the angular scales considered.
However, polarized thermal dust dominates the $B$-mode power on angular scales below
multipole $\ell$ of 200~\citep{bk15}.
\cite{core2017} showed that by including the foreground power spectrum as a noise term in 
the C$^{-1}$ filter of the input CMB fields, the polarization estimator can recover unbiased
lensing amplitude to sub-percent accuracy.
In this work, Galactic dust power in $EE$ and $BB$ at the same level as the simulation inputs 
are included as noise terms in the map filters.
Finally, if Galactic foregrounds had been a significant source of bias, we would fail the
curl null tests. 
The data pass the curl null test for the \tt, \pp\, and \mv\ estimators, suggesting that 
biases from Galactic foregrounds are subdominant at present. \\

We summarize the systematic uncertainties contributed from each source addressed
in this section in~\reftab{sys_uncer}. 
We add the differences in lensing amplitude from each source in quadrature and include the result as a systematic uncertainty
in our lensing amplitude measurements. 
The total systematic uncertainties are $A_{MV}^{\rm sys} = \mvAmpSys$, 
$A_{\rm \pp}^{\rm sys} = \ppAmpSys$, and $A_{\rm \tt}^{\rm sys} = \ttAmpSys$ respectively. 
The \tt\ systematic uncertainty is subdominant given the statistical uncertainties of the \tt\ lensing amplitude,
while those for the \mv\ and \pp\ reach $\sim$0.4\,$\sigma$ of their respective statistical uncertainties.

\section{Conclusions}
\label{sec:conclusion}
In this work, we present a new measurement of the CMB lensing potential
from the 500\,deg$^2$ \sptpol survey.
We use a quadratic estimator to extract the lensing potential $\phi$ from combinations
of temperature and polarization maps. 
Our per-mode measurement has signal-to-noise similar to that in S15 because the noise
levels of the input maps are similar -- we measure $\phi$ with signal-to-noise better than
unity in the multipole range  $100 < L \lesssim 250$. 

We measure the lensing amplitude with our minimum-variance combination of estimators
to be 
$A_{\rm\mv} =  \mvAmp \pm \mvAmpStat {\rm\,(Stat.)}\pm\mvAmpSys{\rm\,(Sys.)}$. 
Of the published lensing measurements using data only from ground-based telescopes, this is
the tightest measurement of the lensing amplitude to date.
Using unlensed simulations to estimate the probability of the measurement, 
we find that in the absence of true CMB lensing our measurement would be 
a \mvNullSig\ outlier -- i.e., we detect lensing at very high 
significance.
We make individual combinations of temperature and polarization estimators. 
Considering statistical uncertainties alone, our polarization-only lensing constraint
is the most precise amongst measurements of its kind to date. 
We perform null tests and consistency checks on our results and find no evidence for 
significant contamination. 
We estimate the size of the systematic uncertainty in our lensing measurement
from uncertainties in calibration, beam measurement, T$\rightarrow$P leakage correction, 
global polarization rotation, and subtraction of extragalactic foreground bias. 
We find that our systematic uncertainty is nearly half as large as our statistical uncertainty 
and dominated by the uncertainty in the polarization calibration. 
Looking forward, we will have to either
improve the precision of the polarization calibration
or find solutions to self-calibrate the lensing estimator in order to reduce 
this systematic uncertainty.

The lensing spectrum measurement in this work is at sufficiently high precision
to provide relevant independent
constraints on cosmological parameters like $\Omega_M$, $\sigma_8$, and the sum of
neutrino masses. 
We will report cosmological parameter constraints in a future paper.
While this work represents the first analysis in which the lensing amplitude is better constrained
by polarization, rather than temperature data, 
this will become typical for future experiments as map noise levels continue to decrease.
The $EB$ estimator will eventually become the most constraining CMB lensing 
estimator, at least for lensing multipoles below several hundred.
For the expected survey map depths of the currently operating SPT-3G experiment
(3.0, 2.2, and 8.8\,$\mu$K-arcmin at 95, 150, and 220\,GHz respectively, ~\citealt{bender18}), the 
$EB$ estimator is projected to provide the highest signal-to-noise lensing amplitude measurement of 
all pairs of input CMB maps. 

The advantages of using polarization maps for CMB lensing measurements are as follows.
First, foregrounds in CMB temperature maps, e.g. tSZ, are the main sources of bias 
in cross correlations with other dark matter tracers~\citep{desy1-5x2, namikawa19}.
Polarization maps are less contaminated by extragalactic foregrounds, and thus more
modes on small angular scales (high $\ell$) can be used for lensing reconstruction and for
cross correlation,
improving the S/N of the measurement in addition to reducing foreground biases.
Second, lensing estimators are typically limited by the \nzero\ noise contributed by the
unlensed power in the input maps. 
The $EB$ estimator, however, is not as limited because unlensed $B$ power is 
subdominant to lensing $B$ power on subdegree scales. 
Furthermore, to first order in $\phi$, lensing $B$ modes are sourced completely by $E$ modes.
These factors 
make methods beyond the quadratic estimator that approach the optimal solution~\citep{millea17} 
particularly efficient in improving the
S/N of the $EB$ estimator. 

High-S/N lensing measurements are essential for constraining the sum of neutrino
masses and the amplitude of primordial gravitational waves, parameterized by the tensor-to-scalar ratio $r$. 
In the coming few years, the uncertainty on $r$ from upcoming CMB experiments will be 
limited by the variance of the lensing $B$ modes. 
To make progress in constraining $r$, we need lensing measurements with high S/N per mode
for delensing~\citep{manzotti17}.
For example, 
delensing with the lensing map from SPT-3G will be crucial for the BICEP Array experiment~\citep{hui18}
to reach its projected uncertainty on $r$ of $\sigma(r) \sim 0.003$, in which 
delensing improves $\sigma(r)$ by a factor of about 2.5. 
For the next-generation ground-based CMB experiment CMB-S4, 
delensing is even more crucial
for achieving the projected $r$ constraint of $\sigma(r) = 5\times10^{-4}$~\citep{cmbs4-sb1}.

\acknowledgements{
The authors would like to acknowledge helpful comments from Chang Feng and Srinivasan Raghunathan
on the manuscript. 
The South Pole Telescope program is supported by the National Science Foundation through grant PLR-1248097.
Partial support is also provided by the NSF Physics Frontier Center grant PHY-0114422 to the Kavli Institute of Cosmological Physics at the University of Chicago, the Kavli Foundation, and the Gordon and Betty Moore Foundation through grant GBMF\#947 to the University of Chicago.  
This work is also supported by the U.S. Department of Energy. 
W.L.K.W is supported in part by the Kavli Institute for Cosmological Physics at the University of Chicago through grant NSF PHY-1125897 and an endowment from the Kavli Foundation and its founder Fred Kavli.
J.W.H. is supported by the National Science Foundation under Award No. AST-1402161.
C.R. acknowledges support from an Australian Research Council Future Fellowship (FT150100074). 
B.B. is supported by the Fermi Research Alliance LLC under contract no. De-AC02-07CH11359 with the U.S. Department of Energy.  
The Cardiff authors acknowledge support from the UK Science and Technologies Facilities Council (STFC).
The CU Boulder group acknowledges support from NSF AST-0956135.  
The McGill authors acknowledge funding from the Natural Sciences and Engineering Research Council of Canada, Canadian Institute for Advanced Research, and the Fonds de Recherche du Qu\'ebec -- Nature et technologies.
The UCLA authors acknowledge support from NSF AST-1716965 and CSSI-1835865.
Work at Argonne National Lab is supported by UChicago Argonne LLC, Operator of Argonne National Laboratory (Argonne). 
Argonne, a U.S. Department of Energy Office of Science Laboratory, is operated under contract no. DE-AC02-06CH11357. 
We also acknowledge support from the Argonne Center for Nanoscale Materials.  
This research used resources of the National Energy Research Scientific Computing Center (NERSC), a U.S. Department of Energy Office of Science User Facility operated under Contract No. DE-AC02-05CH11231.
The data analysis pipeline also uses the scientific python stack \citep{hunter07, jones01, vanDerWalt11} and the HDF5 file format \citep{hdf5}.
G. Longhi's color scheme suggestions are gratefully acknowledged. 
}

\ifdefined\newapj
\bibliographystyle{aasjournal} 
\else
\bibliographystyle{aasjournal_arxiv} 
\fi

\bibliography{spt}

\end{document}